# Mass spectral evidence for iron sulfate in Venus' cloud aerosols

Rakesh Mogul [a,b,*], Mikhail Yu Zolotov [c], Michael J. Way [d,e], Sanjay S. Limaye [f]

[a] Chemistry & Biochemistry Department, California State Polytechnic University, Pomona, CA 91767, USA
[b] Blue Marble Space; Seattle, WA 98104, USA
[c] School of Earth and Space Exploration, Arizona State University; Tempe, AZ 85287, USA
[d] NASA Goddard Institute for Space Studies; New York, NY, 10025, USA
[e] Theoretical Astronomy Department of Physics and Astronomy, Uppsala University; Uppsala, SE-75120, Sweden
[f] University of Wisconsin; Madison, WI 53706, USA

ABSTRACT

Venus' cloud aerosols play pivotal roles in the thermal, chemical, and optical properties of the atmosphere. Here, we present mass spectral evidence that the cloud aerosols contain abundant iron sulfates. Data from the Pioneer Venus Large Probe Neutral Mass Spectrometer (LNMS) suggest that cloud aerosols inadvertently collected into the inlets at ~48–51 km and thermally decomposed below the clouds, releasing several compounds into the LNMS. At ≤23 km, multiple iron species ($^{54}Fe^+$, $^{56}FeO^+$, $^{54}FeO^+$, $^{56}Fe_2O^+$, and $^{56}Fe_2O_2^+$) and other decomposition gases were detected. Quantitative treatment of these mass signals suggested the release of $Fe_2O_3$, $SO_2$, $O_2$, $SO_3$, $H_2SO_4$, and $H_2O$, consistent with the thermal decomposition of hydronium-bearing ferric sulfates. The resulting mole ratio for iron to sulfate (2.0 ± 1.1) aligns with hydronium jarosite ($(H_3O)Fe_3(SO_4)_2(OH)_6$), which likely formed within the inlets after aerosol capture, dehydration, and loss of sulfuric acid. Our thermochemical calculations support these assessments. Speciation of the ferric sulfates in the native aerosols remains unconstrained (*e.g.*, solution and/or particle phases). Peak fits to the thermal release profiles yielded the abundances of the bulk iron (0.8 ± 0.5 mg $m^{-3}$) and aerosol-derived compounds, including hydronium jarosite (2.2 ± 1.2 mg $m^{-3}$, 30 ± 17 wt%), water (3.7 ± 0.4 mg $m^{-3}$, 50 ± 10 wt%), and sulfuric acid (1.5 ± 0.2 mg $m^{-3}$, 20 ± 5 wt%). Surface sourcing of the cloud iron is likely minor. Instead, cosmic dust influx could deliver iron to the clouds in a geologically short time (2.4 ± 1.5 Myr).

## 1. Introduction

The aerosols in Venus' clouds play pivotal roles in the thermal balance, chemistry, and optical properties of the atmosphere, and contribute to the absorption of solar ultraviolet (UV) radiation (Crisp, 1986; Mills *et al.*, 2007; Dai *et al.*, 2025). Most studies suggest that the aerosol composition is dominated by concentrated sulfuric acid ($H_2SO_4$) (≥75%, by mass) (*e.g.*, Dai *et al.* (2025)). However, *in situ* X-ray radiometry data from the Venera 12 and Vega 2 probes suggested the potential presence of chlorine and iron in the cloud aerosols (Petryanov *et al.*, 1981; Andreichikov *et al.*, 1987). This is noteworthy since iron chloride and sulfate compounds are UV absorbers (Bastian *et al.*, 1953) and can contribute to the atmospheric redox chemistry (Frankland *et al.*, 2017).

For Venus' cloud tops, several lines of evidence suggest that the purported unknown UV absorbers are iron compounds. Kuiper and Sill (1969) proposed ferrous chloride ($FeCl_2$) as a major UV absorber between 200–400 nm. Zasova *et al.* (1981) considered ferric chloride ($FeCl_3$) as a minor aerosol species based on matches to the UV albedo (320–450 nm). Markiewicz *et al.* (2014) and Petrova (2018) showed reasonable matches between Venus' glory (365 nm) and ferric chloride (in concentrated sulfuric acid) based on the index of refraction. Jiang *et al.* (2024) demonstrated absorbance matches between the cloud tops (300–500 nm) and mixtures of solid ferric sulfates (rhomboclase and acid ferric sulfate). Egan *et al.* (2025) also noted matches between the cloud top absorption (300–500 nm) and solutions of ferric chloride (in hydrochloric acid or mixtures of hydrochloric acid and sulfuric acid).

Recently, we reported the qualitative detection of iron in the cloud aerosols through reanalysis of mass spectral data acquired by the Large Probe Neutral Mass Spectrometer (LNMS) (Mogul *et al.*, 2025). The LNMS was an instrument onboard the Pioneer Venus Large Probe (PVLP), which descended through Venus' atmosphere in 1978 (Hoffman et al., 1980). During the descent, cloud aerosols from ∼51–48 km were inadvertently collected into the LNMS intake inlets. In the hot sub-cloud atmosphere (≤47 km, ∼370–740 K), these captured aerosols thermally decomposed, releasing gases and compounds into the LNMS. In Mogul et al. (2025), the concomitant releases of $SO_2$, $H_2O$, and $O_2$ across two temperature ranges, including the detection of $FeO^+$ at the higher temperatures, were shown to be consistent with the presence of at least two thermally unstable sulfate-bearing species in the captured aerosols, including sulfuric acid and potentially hydrated ferric sulfates.

* Corresponding author at: Chemistry & Biochemistry Department, California State Polytechnic University, Pomona, CA 91767, USA.
*E-mail address:* rmogul@cpp.edu (R. Mogul).

Here, we quantitatively analyze the signals for iron-bearing species in the LNMS data and present the strongest evidence to date for iron sulfate as a major aerosol constituent. We additionally present a new analysis of the associated signals ($SO_2$, $SO_3$, $O_2$, $H_2O$, and $H_2SO_4$) from Mogul et al. (2025). We then describe the acquisition of the iron to sulfate mole ratio from the captured aerosols, assignment of the parent iron species, and calculation of the aerosol mass loading. Lastly, we evaluate potential surface sources of the cloud iron and suggest cosmic dust as a plausible source.

## 2. Methods

### 2.1. Large probe neutral mass spectrometer

The LNMS was a single-focusing, magnetic sector, and electron ionization mass spectrometer, which measured ion counts at 232 pre-selected mass positions, spaced over irregular intervals across 1–208 u (or *m/z* 1–208) (Hoffman et al., 1980; Mogul et al., 2023; Belkhodja et al., 2025). Consequently, several mass positions were skipped and not measured by the preprogrammed sequence. Mass spectra were acquired at 70, 30, and 22 eV. Background spectra were obtained just prior to sampling. During sampling, 45 spectra were acquired between 64.2 and 0.2 km, per the altitude scale in Hoffman et al. (1980). Between 64.2 and 0.2 km, the LNMS inlets were heated at ≤30 K above the atmospheric temperatures (∼245 to 734 K) to yield an approximate range of ∼272 to 735 K (Mogul et al., 2025). The approximate inlet heating rates ranged from ∼20 to 10 K $min^{-1}$ at ∼47–20 km, and ∼10 to 5 K $min^{-1}$ at ≤23 km (Mogul et al., 2025).

### 2.2. Identification of iron-bearing compounds

The LNMS data were parsed for spikes in the ion counts for iron-bearing species. Chemical assignments were based upon (**a**) sufficient overlap between the LNMS mass position and the expected mass peak for the target iron species (Section 2.3), (**b**) the potential to deconvolute the target species from possible isobars, (**c**) measurement of ≥1 count at ≥3 consecutive altitudes in the descent profile, and (**d**) precedence of the chemical species in the literature or NIST database (Wallace, 2020). Under these constraints, the iron-related assignments and associated LNMS mass positions (exact masses in parentheses) were: (**a**) $^{54}Fe^+$ at 54.004 u (53.940 u), (**b**) $^{54}FeO^+$ at 69.971 u (69.935 u), (**c**) $^{56}FeO^+$ at 71.973 u (71.930 u), (**d**) $Fe_2O^+$ at 127.906 u (127.865 u), and (**e**) $Fe_2O_2^+$ at 143.946 u (143.860 u). These iron species were observed only in the sub-cloud atmosphere after aerosol capture (≤23 km; >560 K, inlet temperatures).

### 2.3. LNMS peak fitting

The mass data were treated using the peak-fitting and normalization strategies described in Mogul et al. (2023) and Mogul et al. (2025). Peak fits were conducted for $^{54}Fe^+$, $^{54}FeO^+$, $^{56}FeO^+$, $Fe_2O^+$, $Fe_2O_2^+$, $SO_2^+$, $SO_3^+$, and $H_2SO_4^+$. These species were represented by 1 mass point each in the data. The LNMS mass positions for the iron-bearing species (Section 2.2) did not match the exact masses; instead, they represented points on the respective peak slopes. The LNMS mass position for $H_2SO_4^+$ (97.948 u) closely matched the exact mass (97.968 u). The LNMS mass positions for $SO_2^+$ and $SO_3^+$ matched the exact masses (*e.g.*, $SO_2^+$ at 63.962 u), presumably because these species were pre-selected for analysis. Given these peak representations, the peak centers were defined as the exact mass. The expected peak widths (*e.g.*, full width half maximum, FWHM) were obtained by interpolation ($^{54}Fe^+$, $^{54}FeO^+$, $^{56}FeO^+$, $SO_2^+$, $SO_3^+$, $H_2SO_4^+$, and $Fe_2O^+$) or extrapolation ($Fe_2O_2^+$) using the linear relationship (at each altitude) between the FWHM and exact mass of the LNMS internal standards. The chosen LNMS internal standards were $CH_3^+$, $CO^+$, $N_2^+$, $^{40}Ar^+$, and $^{136}Xe^+$ (averaged $R^2 = 0.999 \pm 0.001$; 22 spectra, 70 eV; ≤24 km).

### 2.4. Isotopic composition of chlorine

Deconvolution of $FeO^+$ from $Cl_2^+$ required the $^{35}Cl/^{37}Cl$ ratio. The $^{35}Cl/^{37}Cl$ ratio was obtained using the counts for $^{35}Cl^+$ and $^{37}Cl^+$ from the LNMS cloud data (58.3–50.3 km). The multi-step deconvolution procedure, described in **Appendix A**, was refined from Mogul et al. (2021) and used the peak-fitting procedures from Mogul et al. (2025).

### 2.5. Disambiguating iron from isobars

Relevant isobars were deconvoluted from $^{56}FeO^+$, $^{54}FeO^+$, and $^{56}Fe_2O^+$ (LNMS mass positions are in Section 2.2). At 72 u, $^{35}Cl^{37}Cl^+$ (71.935 u, exact mass) was treated as a minor isobar to $^{56}FeO^+$ (71.930 u, exact mass). At 70 u, $^{35}Cl_2^+$ (69.938 u, exact mass), referred to herein as $Cl_2^+$, was treated as a major isobar to $^{54}FeO^+$ (69.935 u, exact mass). These isobars were deconvoluted algebraically (**Eqs. 1-4**) by substitution using the $^{35}Cl/^{37}Cl$ and $^{56}Fe/^{54}Fe$ ratios. The $^{54}Fe/^{56}Fe$ ratio (0.0637) was from Brand et al. (2014). The $^{35}Cl/^{37}Cl$ ratio was obtained in Section 2.4.

$$I_{72\ u} = \ ^{56}FeO^+ + ^{35}Cl^{37}Cl^+ \tag{1}$$

$$I_{70\ u} = \ ^{54}FeO^+ + Cl_2^+ \tag{2}$$

$$I_{70\ u} = \ ^{56}FeO^+\left(\frac{^{54}Fe}{^{56}Fe}\right) + ^{35}Cl^{37}Cl^+\left(\frac{^{35}Cl}{2^{37}Cl}\right) \tag{3}$$

$$^{56}FeO^+ = \frac{I_{70\ u} - \left(I_{72\ u}\left(\frac{^{35}Cl}{2^{37}Cl}\right)\right)}{\left(\left(\frac{^{54}Fe}{^{56}Fe}\right) - \left(\frac{^{35}Cl}{2^{37}Cl}\right)\right)} \tag{4}$$

In **Eq.** (1), the total ion counts at 72 u ($I_{72\ u}$) were defined as the sum of $^{56}FeO^+$ and $^{35}Cl^{37}Cl^+$. In **Eq.** (2), the total ion counts at 70 u ($I_{70\ u}$) were defined as the sum of $^{54}FeO^+$ and $Cl_2^+$. To yield **Eq.** (3), the following relationships were substituted into **Eq.** (2): (**a**) $^{54}FeO^+ = {}^{56}FeO^+(^{54}Fe/^{56}Fe)$ and (**b**) $Cl_2^+ = (^{35}Cl^{37}Cl^+)(^{35}Cl/(2^{37}Cl))$. Counts for $^{56}FeO^+$ at 72 u were thus obtained by (**a**) rearranging **Eq.** (1) to $^{35}Cl^{37}Cl^+ = I_{72\ u} - \ ^{56}FeO^+$, (**b**) substituting this expression into **Eq.** (3), (**c**) rearranging to yield **Eq.** (4), and (**d**) solving for $^{56}FeO^+$. All uncertainties were propagated.

The isobaric species at 128 u ($^{56}Fe_2O^+$ and $^{128}Xe^+$) were deconvoluted by subtracting the expected counts for $^{128}Xe^+$ at each altitude. The $^{128}Xe^+$ counts arose from the onboard LNMS calibration gas mixture, which was composed of $CH_4$, $^{136}Xe$, and several trace xenon isotopes (Mogul et al., 2023). Expected counts for $^{128}Xe^+$ were calculated by multiplying the fitted counts for $^{136}Xe^+$ (Mogul et al., 2023) by the $^{128}Xe^+$ percent abundance from the LNMS background spectra (0.035 ± 0.012%, relative to $^{136}Xe^+$).

### 2.6. Reaction pathways and stoichiometry of ferric sulfate decomposition

The thermal decomposition for mikasaite ($Fe_2(SO_4)_3$) and/or hydronium jarosite ($(H_3O)Fe_3(SO_4)_2(OH)_6$) in the LNMS inlets were assumed to proceed through Reactions (5-9), consistent with laboratory studies on mineral dehydration and decomposition (Šolc et al., 1988;

Frost *et al.*, 2006; Spratt *et al.*, 2014). In Reaction (5), mikasaite decomposes to hematite (α-$Fe_2O_3$) and $SO_3$ gas. Reaction (6) is a potential dehydroxylation pathway for hydronium jarosite, resulting in the loss of $H_2O$. Based on reported experimental data (Šolc et al., 1988; Frost et al., 2006), basic ferric sulfate ($FeOHSO_4$) and/or similar compounds (*e.g.*, $Fe_2O(SO_4)_2$) were assumed to be intermediates in the decomposition of hydronium jarosite. Reaction (7) illustrates the final dehydroxylation of $FeOHSO_4$. Reaction (8) is the overall thermal decomposition pathway for hydronium jarosite. In Reaction (9), $SO_3$ from Reactions (5), (7), and/or (8) decomposes to $SO_2$ and $O_2$.

$$Fe_2(SO_4)_3 \rightarrow Fe_2O_3 + 3SO_3 \quad (5)$$

$$(H_3O)Fe_3(SO_4)_2(OH)_6 \rightarrow 2FeOHSO_4 + 0.5Fe_2O_3 + 3.5H_2O \quad (6)$$

$$2FeOHSO_4 \rightarrow Fe_2O_3 + H_2O + 2SO_3 \quad (7)$$

$$(H_3O)Fe_3(SO_4)_2(OH)_6 \rightarrow 1.5Fe_2O_3 + 4.5H_2O + 2SO_3 \quad (8)$$

$$SO_3 \rightarrow SO_2 + 0.5O_2 \quad (9)$$

The stepwise electron ionization and fragmentation of $Fe_2O_3$ particles in the LNMS were assumed to proceed through Reactions (10) and (11). In Reaction (10), electron ionization (70 eV) of $Fe_2O_3$ yields $Fe_2O_3^+$, consistent with the cross sections in Huber *et al.* (2017). In Reaction (11), fragmentation of $Fe_2O_3^+$ yields $Fe^+$, $FeO^+$, $Fe_2O_2^+$, and other species (Li *et al.*, 2009; Yan *et al.*, 2010).

$$Fe_2O_3 + e^- \rightarrow Fe_2O_3^+ + \text{other products} \quad (10)$$

$$Fe_2O_3^+ \rightarrow Fe^+ + FeO^+ + Fe_2O_2^+ + \text{other products} \quad (11)$$

### 2.7. LNMS aerosol composition

Molecular abundances ($N$), or the number of molecules released from the captured aerosols during the thermal decomposition, and the associated uncertainties were extracted from the LNMS data (Mogul *et al.*, 2025). The molecular abundances of $^{56}Fe_2O_3$ (Reactions (5, 7, and/or 8)) were obtained using the $^{56}FeO^+$ counts as a proxy. The sulfate molecular abundances were obtained using the $SO_2^+$ counts as a proxy (Reaction (9)). Ion counts ($I_i$) for the target species ($i$) were converted to $N_i$ through **Eq.** (12) using a species-specific sensitivity factor ($\rho_i$). The target-specific $\rho_i$ was related through **Eq.** (13) to the behavior of $CO_2$ (the dominant species in the LNMS data) using the respective ratios of the cross sections for the parent species ($\sigma_i$), fraction of the target species among the measured fragment ions ($f_i$), transmission coefficients for the target species ($\nu_i$), and the exact molecular mass of the target species ($MW_i$).

$$N_i = \rho_i I_i \quad (12)$$

$$\rho_i = I_i\left(\frac{\sigma_{CO_2}}{\sigma_i}\right)\left(\frac{f_{CO_2}}{f_i}\right)\left(\frac{\nu_{CO_2}}{\nu_i}\right)\left(\frac{MW_{CO_2}}{MW_i}\right)^{1/2} \quad (13)$$

The cross section (70 eV) for the electron ionization of $Fe_2O_3$ (Reaction (10)) was obtained by averaging the calculated values in Huber *et al.* (2017) ($\sigma = 8.5\times10^{-16} \pm 1.5\times10^{-16}$ cm$^2$; ~18% uncertainty). The fraction of $^{56}FeO^+$ ($f_i$) produced by the fragmentation of $^{56}Fe_2O_3^+$ was calculated using the LNMS mass data and **Eq.** (14).

$$f = {}^{56}FeO^+/({}^{56}Fe^+ + {}^{56}FeO^+ + {}^{56}Fe_2O^+ + {}^{56}Fe_2O_2^+) \quad (14)$$

For **Eq.** (14), the counts for $^{56}Fe^+$ were unavailable since the LNMS mass sweep skipped measurements at 56 u (Sections 2.1). Therefore, the ion counts for $^{56}Fe^+$ were estimated by multiplying the counts of $^{54}Fe^+$ by the expected $^{56}Fe/^{54}Fe$ ratio of 15.7 (≤0.19% uncertainty), consistent with the iron isotopic composition in chondritic ($\delta^{56}Fe$ of -1.9 to +0.5 ‰) and terrestrial ($\delta^{56}Fe$ of ±0.10 ‰) samples (Beard and Johnson, 2004; Brand *et al.*, 2014). The corresponding $f_i$ values (11.1–23.0 km) were $^{56}Fe^+$ (97.7 ± 1.1%), $^{56}FeO^+$ (2.23 ± 1.06%), $^{56}Fe_2O^+$ (0.08 ± 0.02%), and $^{56}Fe_2O_2^+$ (0.08 ± 0.07%). At 30 and 22 eV, no counts for $^{56}FeO^+$ were measured. The $\nu_i$ terms were obtained as described in Mogul *et al.* (2025). The σ and $f_i$ terms for $SO_3^+$ and $H_2SO_4^+$ were estimated from Snow and Thomas (1990) and Parikh et al. (2022), while those for $SO_2^+$ and $CO_2^+$ were obtained as detailed in Mogul *et al.* (2025). All uncertainties were propagated.

### 2.8. LNMS thermal evolution profiles

Thermal evolution profiles for the captured aerosols were constructed by plotting $N_i$ for the decomposition products ($Fe_2O_3$, $SO_2$, $SO_3$, and $H_2SO_4$) against the LNMS inlet temperatures at the measurement altitudes (Section 2.1). The thermal profile for $SO_2$ included measurements obtained at 70, 30, and 22 eV. As described (Mogul *et al.*, 2025), the thermal profiles were fitted to Gauss functions and the regressions minimized by least squares analysis, which yielded the temperatures at maximum decomposition (peak center; $T_D$, K), maximum number of released molecules (peak height; $N_i^{max}$), and the associated standard deviation for $N_i^{max}$ ($\sigma_i^{stdev}$). The inclusion of multiple sequential and overlapping peaks yielded excellent approximations to the thermal profiles. The peak areas represented the total number of molecules released at each decomposition step.

### 2.9. Relative abundances and mass loading

Molecular abundances and mass fractions for the parent compounds were obtained using the stoichiometric relationships in Reactions (5–10). The $N_i$ for $^{56}Fe_2O_3$, referred to herein as $Fe_2O_3$, was converted to the respective $N_i$ for $Fe^{3+}$, $Fe_2(SO_4)_3$, or $(H_3O)Fe_3(SO_4)_2(OH)_6$. The $N_i$ for $SO_2$ was converted to the $N_i$ for sulfate ($SO_4^{2-}$). Relative abundances were expressed as weight percent (wt%) of the thermally unstable and volatile compounds in the captured aerosols. Water abundances were adjusted to account for the thermal decomposition stoichiometry of hydronium jarosite (Reaction (8)) and $H_2SO_4$ (Mogul *et al.*, 2025). As detailed in Mogul *et al.* (2025), mass loading values were estimated using an aerosol collection column with a diameter defined by the rapidly spinning PVLP probe (~0.98 m) and a column height (2.6 km) assigned as 48.4 to 51.0 km, which correlated to where the inlet clog began (~51.0 km) and $CO_2^+$ counts decreased by 95% (48.4 km), indicating near stoppage of intake.

### 2.10. Geochemical residence time

The geochemical residence time ($\tau$) for the cloud iron was calculated using **Eq.** (15). Relevant terms included the (**a**) influx rate for ablated iron at Venus ($\phi = 1.5\times10^6 \pm 0.9\times10^6$ kg yr$^{-1}$) (Carrillo-Sánchez *et al.*, 2020), (**b**) total cloud iron mass ($m_{Fe}$), (**c**) mass percent of iron (wt%), (**d**) aerosol mass loading ($M$, kg m$^{-3}$) between $i = 47$ km and $k = 51$ km, and (**e**) shell volume ($V$, m$^3$) between 47 and 51 km, which correlated to where the capture of clog-inducing aerosols by the LNMS began (~51 km) and the cloud deck ends (~47 km). **Appendix B** details the acquisition of the Fe and sulfate masses (kg) from the ferric sulfates at ~47–51 km.

$$\tau = m_{Fe} \Big/ \phi = \left(\sum_i^k (wt\%) M_i V_i\right) \Big/ \phi \quad (15)$$

### 2.11. Stability assessments for iron compounds

Thermal stabilities were assessed by calculating the chemical equilibria conditions between iron-bearing solids (mikasaite, hydronium jarosite, $FeSO_4$, hematite, magnetite, pyrrhotite, and pyrite) and atmospheric gases ($SO_3$, $SO_2$, $H_2O$, CO, and $CO_2$). Stabilities were assessed under the sub-cloud atmosphere conditions (Seiff *et al.*, 1985), within the LNMS inlet assembly (Mogul *et al.*, 2025), and at the surface. Stability of mikasaite at sub-cloud atmospheric temperatures was evaluated by calculating the partial pressure ($p$) of $SO_3$ at the mikasaite-hematite equilibrium (Reaction (5)) using the mikasaite thermodynamic data from Majzlan *et al.* (2005). The stability of hydronium jarosite was estimated by calculating $pSO_3$ (Reaction (8)) using data from Majzlan *et al.* (2004). Mineral stabilities within the LNMS inlets after aerosol capture were assessed using the *apparent* mixing ratios for the evolved water vapor within the inlets from Mogul *et al.* (2025). Thermal decomposition profiles for other metal sulfates (e.g., $NiSO_4$, $MgSO_4$, $CaSO_4$) and their reaction products (metal oxides and $SO_3$) were calculated. The chemical stabilities for iron sulfates, sulfides, and oxides under the conditions of Venus' modal radius surface (740 K, 95.6 bar, 6051.4 km) were estimated as a function of $pSO_2$ and $pCO$ at $pCO_2$ (92.3 bar; 96.5 vol%).

The stabilities of ferric chlorides relative to ferric oxides at the surface were also evaluated. Calculations included the chloride-oxide phase equilibria with solid or gaseous iron chlorides (*e.g.*, $FeCl_2$, $FeCl_3$, $Fe_2Cl_6$, and $FeF_3$), as exemplified by Reaction (16).

$$Fe_2O_{3(s)} + 4HCl_{(g)} + CO_{(g)} \rightarrow 2FeCl_{2(g)} + 2H_2O_{(g)} + CO_{2(g)} \quad (16)$$

To estimate speciation of the Fe-bearing gases, gas-phase chemical equilibria were calculated in the C-O-N-S-H-Cl-F-Fe system at the conditions of the modal radius. The atmospheric composition included the following gases: 96.5 vol% $CO_2$, 3.5 vol% $N_2$, 150 ppm $SO_2$, 30 ppm $H_2O$, 17 ppm CO, 0.4 ppm HCl, and 5 ppb HF. The Fe/C atomic ratio was treated as a free parameter. Thermodynamic data and uncertainties for metal sulfates, sulfides, and oxides, as well as gases, were mostly from Robie and Hemingway (1995).

## 3. Results

### 3.1. Peak fits and chemical assignments

The iron-bearing compounds inferred from the LNMS data included $^{54}Fe^+$, $^{56}FeO^+$, $^{54}FeO^+$, $^{56}Fe_2O^+$, $^{56}Fe_2O_2^+$. Very reasonable peak fits to the mass data were obtained for the iron-bearing species, $SO_2^+$, $SO_3^+$, and $H_2SO_4^+$ (Fig. 1). Comparison of the fitted counts for $^{56}FeO^+$ and $^{54}FeO^+$ yielded an improbably low $^{56}Fe/^{54}Fe$ isotope ratio (4.6 ± 3.0), indicating the presence of isobars, presumably $^{35}Cl^{37}Cl^+$ and $Cl_2^+$, respectively. Deconvolution of these isobars required the Venus ratio for $^{37}Cl/^{35}Cl$ (0.35 ± 0.03), which was obtained from the LNMS cloud data (**Appendix A**). The LNMS $^{37}Cl/^{35}Cl$ ratio is comparable to the terrestrial seawater value (0.31970) (Magenheim *et al.*, 1994), when considering the error and uncertainties from the assumptions (**Appendix A**). Relative to the LNMS $^{37}Cl/^{35}Cl$ ratio, the seawater $^{37}Cl/^{35}Cl$ ratio yielded ~1% differences in the $^{56}FeO^+$ counts.

The LNMS signals for $^{54}Fe^+$, $^{56}FeO^+$, $^{54}FeO^+$, $^{56}Fe_2O^+$, $^{56}Fe_2O_2^+$ were assigned as the electron ionization products of $Fe_2O_3$ (Reactions (10) and (11)), consistent with the limited fragmentation studies regarding ferric oxides (Li *et al.*, 2009; Yan *et al.*, 2010). For example, Li *et al.* (2009) showed that the collision-induced dissociation reactions between $Fe_2O_x^+$ ($x = 1$–3) and Xe yielded $Fe^+$, $FeO^+$, $Fe_2O^+$, $Fe_2O_2^+$, and other products.

Iron abundance in the captured aerosols was obtained from the $^{56}FeO^+$ counts. The thermal release steps for $^{56}FeO^+$, when compared to $^{54}Fe^+$, better correlated with the $SO_2$ spikes. This was likely due to unaccountable isobars to $^{54}Fe^+$, predominantly at <11 km, such as $C_4H_6^+$ isomers. Appreciable uncertainty was observed for the $f_{^{56}FeO^+}$ term (47%), which potentially indicated (**a**) unaccountable mass errors and/or (**b**) the release of $Fe_2O_3$ and other ferric oxy (or, Fe(III)O-bearing) species possessing differing ionization and fragmentation yields. The $f_{^{56}FeO^+}$ uncertainty was propagated to account for either possibility.

Robust chemical assignments for other iron-bearing species were not possible. The assignment of $^{56}Fe^+$ (55.935 u) was impossible since the 56 u mass position was skipped and not measured during the LNMS measurement sequence. The assignment of $FeO_3^+$ (103.920 u), despite sufficient counts at 103.985 u (≤11 counts), was considered implausible since $FeO_3^+$ is a product of high-energy irradiations (Yan *et al.*, 2010) and collision-induced dissociation of $FeO_4^+$ (Li *et al.*, 2009). The assignment for $Fe_2O_3^+$ (159.855 u) is putative due to 1 count being measured (at 159.859 u) at 4 non-consecutive altitudes (0.9–19.1 km).

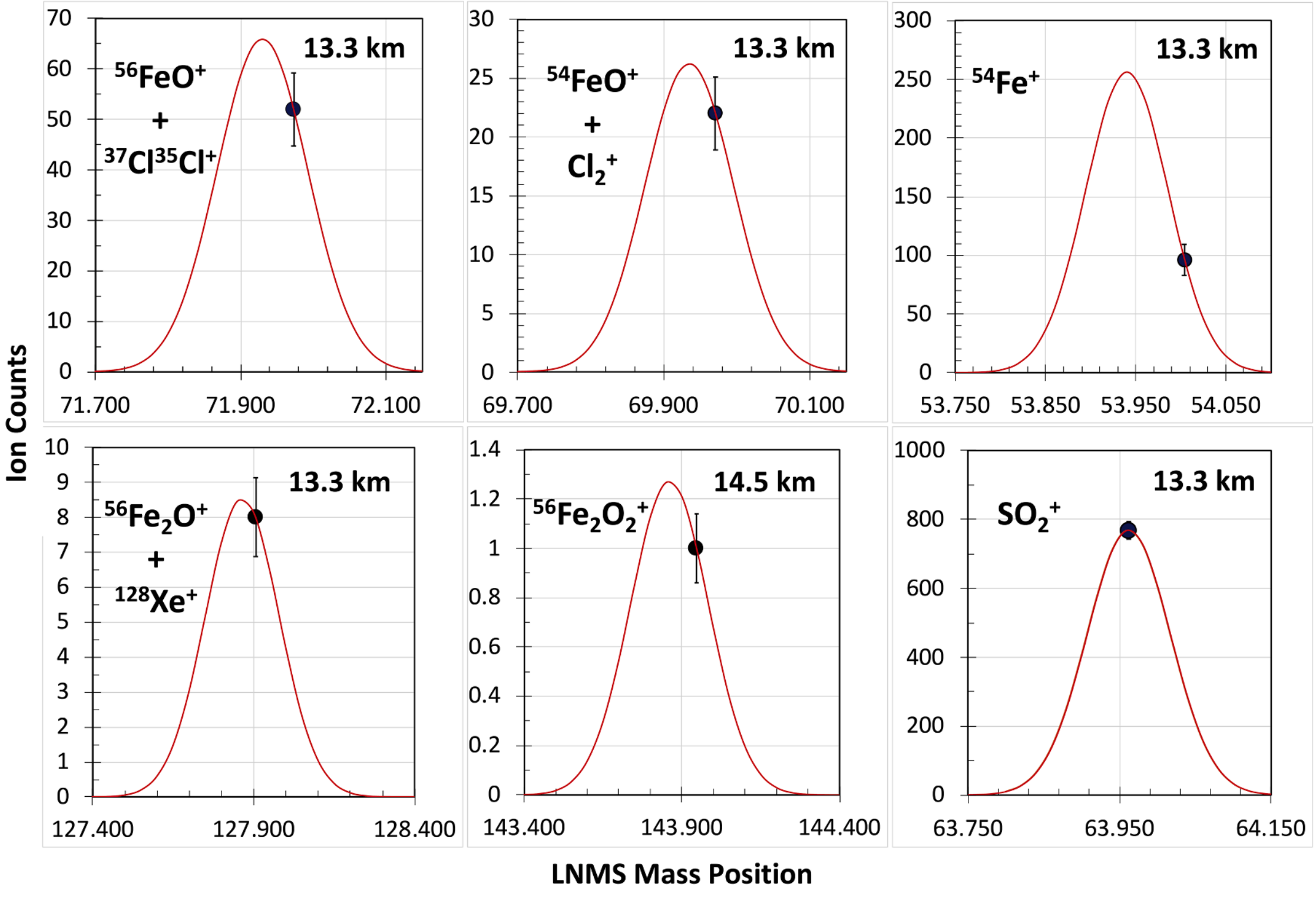

**Fig. 1.** Example peak fits to the LNMS mass positions. Respective species at 72, 70, 54, 128, 144, and 64 u are indicated in the plots.

### 3.2. Thermal decomposition within the LNMS inlet assembly

During the PVLP descent, cloud aerosols collected into the LNMS inlets at ~48–51 km and thermally decomposed across two major temperature ranges (Fig. 2). From ~47–20 km (inlet temperatures of ~380–585 K), sulfuric acid decomposed (Mogul *et al.*, 2025). Then, from ~23 km to the surface (inlet temperatures of ~560–735 K), several other aerosol-derived species released into the LNMS, including $Fe_2O_3$, $SO_2$, $O_2$, $SO_3$, $H_2SO_4$, and $H_2O$ (Figs. 3 and 4). In this study, we improved the profiles for $SO_2$, $O_2$, $SO_3$, $H_2SO_4$, and $H_2O$ from Mogul *et al.* (2025) by (**a**) peak fitting the mass data for $SO_2$, $SO_3$, and $H_2SO_4$ at each altitude, (**b**) peak fitting the $SO_2$ thermal evolution profile using multiple release steps, compared to the one release step in Mogul *et al.* (2025), and (**c**) assuming an uncertainty of 30 ± 5 eV for the spectrum obtained at ~9 km, which yielded more plausible peak profiles for $SO_2$ and $O_2$, and minor changes for $H_2O$.

Treatment of the LNMS data as an evolved gas analysis permitted deconvolution of a stepwise thermal decomposition process. The fitted thermal profile for $Fe_2O_3$ (and/or other Fe(III)O-bearing compounds) suggested ~5 release steps ($T_D$) at 628, 683, 706, 719, and 732 K (Fig. 3a). The fitted thermal profile for $SO_2$ (Fig. 3b), after the aerosol-phase sulfuric acid decomposed (see Fig. 3b legend), suggested ~4 release steps at 593, 640, 676, and 703 K. Compared to $SO_2$, minor abundances were measured for $H_2SO_4$ and $SO_3$ (**Appendix** Fig. C.1), which aligns with their expected and limited ability to pass through the LNMS inlet (Hoffman *et al.*, 1980; Mogul *et al.*, 2025). The $H_2SO_4$ released in 3 broad steps, whereas $SO_3$ released in 5 steps (see **Appendix** Fig. C.1 legend). Dissimilar profiles between $H_2SO_4$ and $SO_3$ were observed at ~600–650 K. At ≥600 and ≥625 K, the $H_2SO_4$ and $SO_3$ abundances increased by ≤10-fold, respectively. At ≤23 km, identical to Mogul *et al.* (2025), $H_2O$ released in a single broad step, while $O_2$ released in ~7 steps.

### 3.3. Aerosol composition: the thermally unstable and volatile constituents

The thermally unstable and volatile aerosol constituents that released $Fe_2O_3$, $SO_2$, and $H_2O$ were treated as hydronium jarosite or mikasaite, sulfuric acid, and water. The respective abundances were obtained using the release profiles for $FeO^+$, $SO_2^+$, and $H_2O^+$ at ≤23 km. Counts were converted using **Eq.** (13) to the molecular abundances of $Fe_2O_3$, $SO_2$, and $H_2O$, and the abundances of $SO_2$ were converted to sulfate using the stoichiometry in Reactions (5) and (9). Listed in Table 1 and **Appendix** Table C.1 are the resulting molecular abundances, mole ratios (iron to sulfate; water to hydronium jarosite, and water to sulfuric acid), relative mass abundances, atmospheric mass loading, and the total aerosol mass loading. Mass loading values were also obtained for the bulk iron, sulfate from the ferric sulfates, and total sulfate from sulfuric acid and ferric sulfates. Irrespective of the ferric sulfate identity, the relative abundances were within error.

### 3.4. Iron geochemical residence time

The geochemical residence time (Broecker and Peng, 1982) was treated as the time required to achieve the cloud iron concentration *via* cosmic dust infall. Estimation of the total iron mass at 47–51 km ($3.5\times10^{12} \pm 1.0\times10^{12}$ kg) is described in **Appendix B**. This procedure accounted for the ~10-fold increases in aerosol mass loading observed between 50–51 and ~49 km (Knollenberg and Hunten, 1980). The iron mass at 47–51 km could be achieved in 2.4 ± 1.5 Myr of cosmic dust infall.

### 3.5. Thermal stabilities under Venus conditions

The thermal decomposition equilibrium conditions for differing metal sulfates were calculated under Venus and LNMS descent conditions (Fig. 5). Hydronium jarosite is stable under the cloud conditions (30 ppm $H_2O$ vapor). However, below the clouds at ~40 km, hydronium jarosite was less thermally stable than mikasaite and hematite, when considering the measured and modeled values of 30 ppm $H_2O$ vapor (Mills *et al.*, 2007) and 0.2 ppm $SO_3$ (Krasnopolsky, 2013), respectively. Within the LNMS inlet, hydronium jarosite would be thermally stable down to 36-37 km, due the *apparent* mixing ratio for the $H_2O$ that released from the captured aerosols within the inlet between ~45–35 km (~2000–7000 ppm; Mogul *et al.* (2025)). Our calculations show that the ferric sulfates were less stable than $CaSO_4$, $MgSO_4$, $Na_2SO_4$, $MnSO_4$, $NiSO_4$, and $CoSO_4$ below the clouds. When considering the $SO_3$ altitude profile (see Fig. 5 legend), mikasaite, $NiSO_4$, and $CoSO_4$ would decompose below ~24–26, 13–15, and 9–11 km, respectively; whereas $CaSO_4$, $MgSO_4$, $Na_2SO_4$, $MnSO_4$ would not decompose.

Equilibrium calculations under surface conditions (Fig. 6) indicate chemical instability of mikasaite relative to ferric oxides ($Fe_xO_y$), when considering the partial pressures of the major chemically active atmospheric gases ($CO_2$, $SO_2$, CO, and OCS). Reactions between mikasaite and CO and/or OCS form hematite and/or magnetite. The oxidation state near the surface (*e.g.*, $O_2$ fugacity and $CO_2$/CO ratio) is

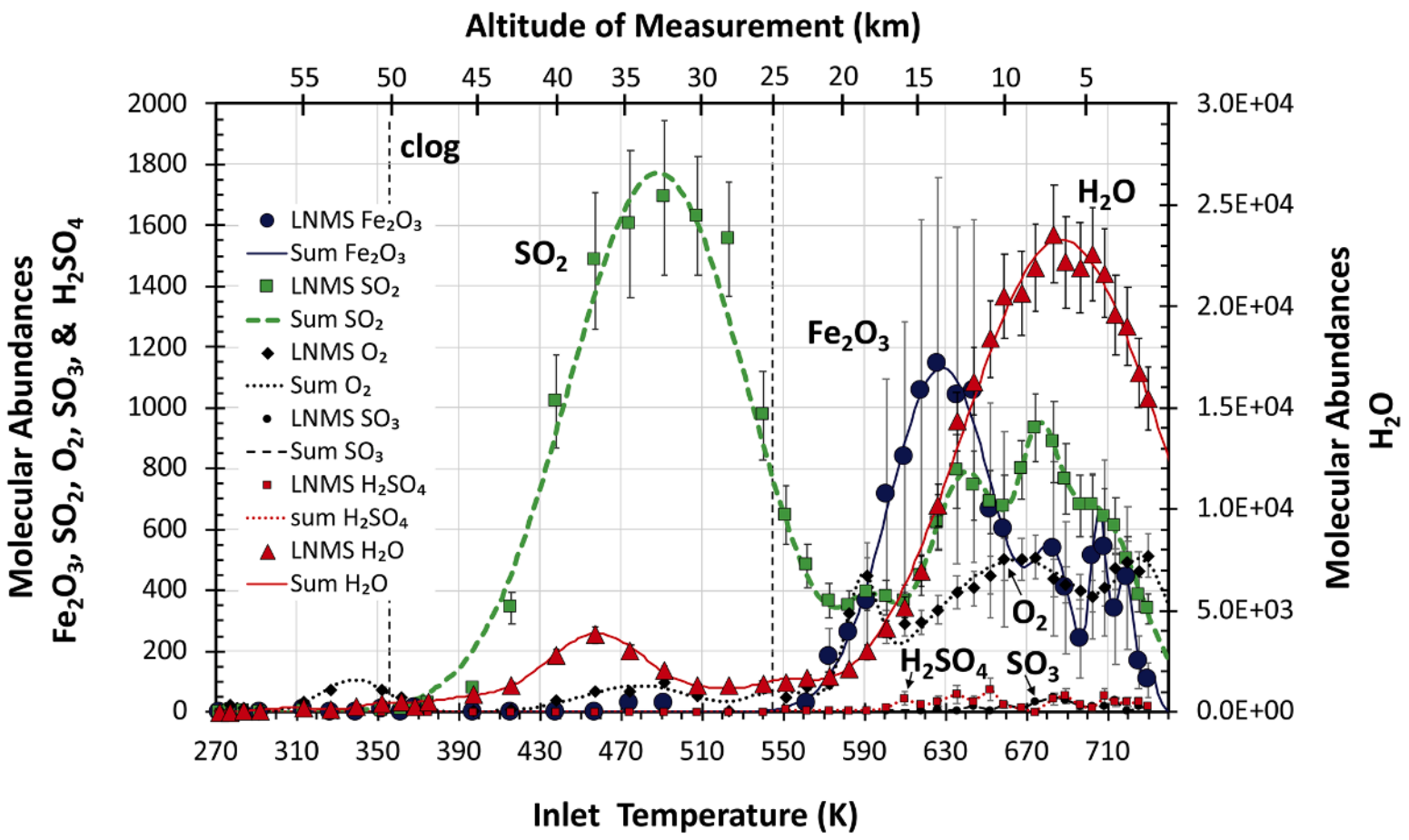

**Fig. 2.** The LNMS thermal evolution profiles and peak fits across the descent profiles (64.2–0.9 km) for $Fe_2O_3$, $SO_2$, $O_2$, $SO_3$, and $H_2SO_4$ (left-hand y-axis) and $H_2O$ (right-hand x-axis). Molecular abundances are plotted against the LNMS inlet temperatures (lower x-axis) and altitudes of measurement (upper x-axis). Fits to the data (solid, dashed, and dotted lines) represent the sum of the Gauss peaks (not shown). The LNMS clog is demarked by the vertical dashed lines at ~50 and 25 km. Propagated errors are displayed.

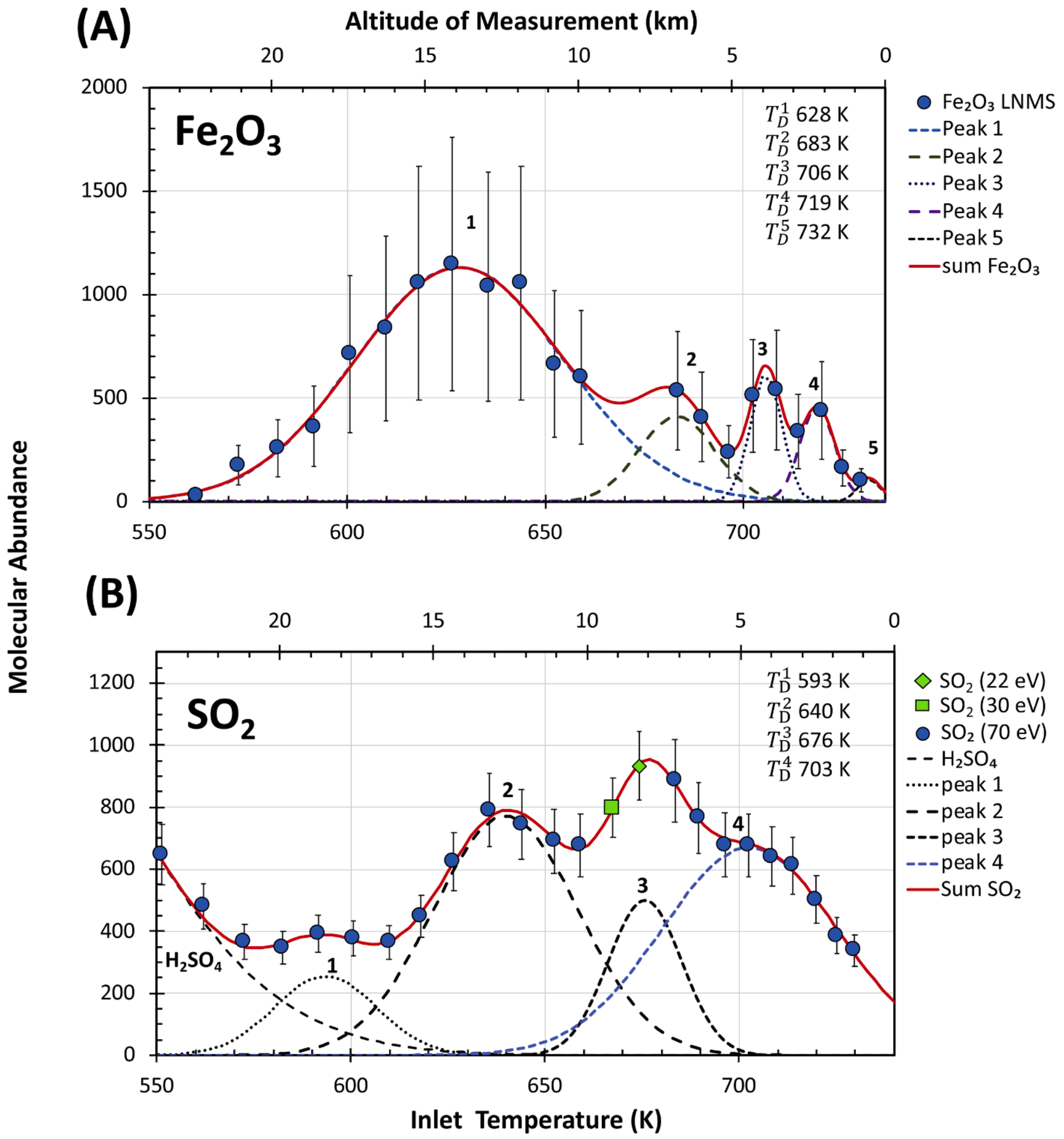

**Fig. 3.** The LNMS thermal evolution profiles and peak fits for (A) $Fe_2O_3$ and (B) $SO_2$. Molecular abundances are plotted against the LNMS inlet temperatures (lower x-axis) and altitudes of measurement (upper x-axis). Fits to the data (solid red lines) represent the sum of the numbered Gauss peaks (dashed and dotted peaks). In (B), the tail end of the $SO_2$ peak associated with the decomposition of the aerosol-phase $H_2SO_4$ is shown. Temperatures at maximum decomposition ($T_D^n$) for each peak (n) are listed. Propagated errors are displayed.

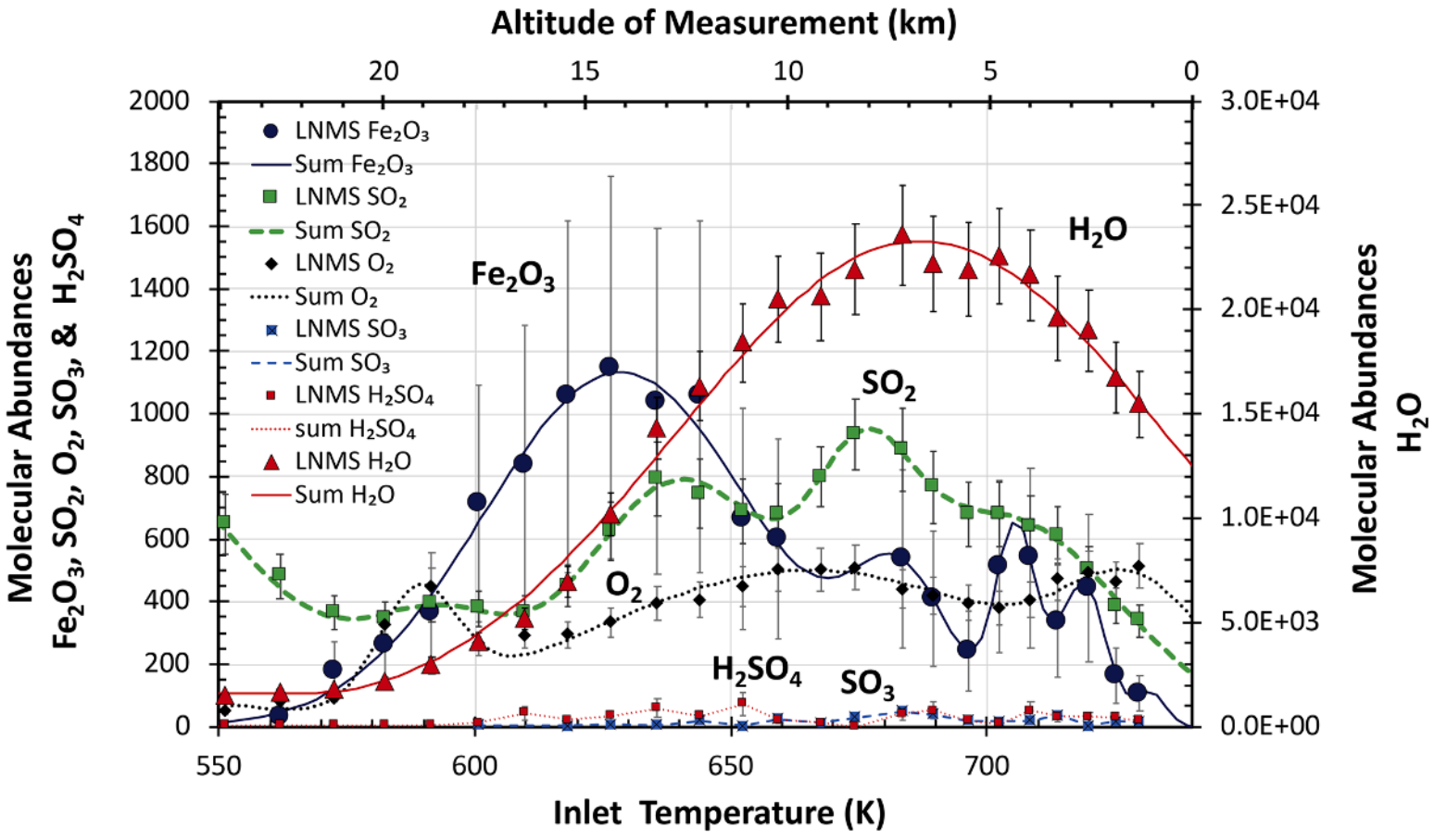

**Fig. 4.** Overlay of the thermal evolution profiles associated with the decomposition of hydronium jarosite (or similar compound), as plotted against the LNMS inlet temperatures (lower x-axis) and altitudes of measurement (upper x-axis). Decomposition products included $Fe_2O_3$, $SO_2$, $O_2$, $SO_3$, and $H_2SO_4$ (left-hand y-axis), and $H_2O$ (right-hand y-axis).

indistinguishable (within error) from the conditions of the hematite-magnetite phase boundary (α-$Fe_2O_3$-$Fe_3O_4$) (Fegley *et al.*, 1997b). This suggests the co-presence of hematite, magnetite, and pyrite at the surface (Zolotov, 2018; Zolotov, 2026). When considering the gradual settling of mikasaite grains toward the surface, hematite could form *via* thermal decomposition (Reactions (5–8)) and slower gas-solid-type reactions (Reactions (17) and (18)). Within the LNMS inlets, hematite likely formed *via* Reactions (5–8) given the rapid descent (~43 min,

**Table 1**
The LNMS aerosol composition and comparisons to Venus values. Aerosols were captured at ~51-48 km. Compositional parameters represent the thermally unstable and volatile aerosol constituents and include the number of released molecules (*N*), weight percentage (wt%), and mass loading (mg $m^{-3}$). The table is organized by (A) bulk abundances for the major aerosol compounds, (B) bulk abundances of selected constituents, and (C) bulk mole ratios. Values for $H_2SO_4$ are from Mogul *et al.* (2025).

| LNMS Aerosol Composition | Venus Values | | | | |
|---|---|---|---|---|---|
| (A) Bulk Aerosol Abundances | | | | | |
| Compounds | N ($\times 10^5$) | wt% | mg $m^{-3}$ | mg $m^{-3}$ | |
| Aerosol $H_2O$ | 26.59 ± 3.05 | 50 ± 10 | 3.7 ± 0.4 | – | |
| ($H_3O$) $Fe_3(SO_4)_{1.5}(OH)_7$ | 0.64± 0.34 | 30 ± 17 | 2.2 ± 1.2 | – | |
| $H_2SO_4$ | 1.95 ± 0.28 | 20 ± 5 | 1.5 ± 0.2 | ~0.4 – 1[a] | |
| Total mass loading | 7.4 ± 1.3 | | | ~12[bb] | |
| (B) Aerosol Constituent Abundances | | | | | |
| Constituents | N ($\times 10^5$) | wt% | mg $m^{-3}$ | mg $m^{-3}$ | |
| Total iron | 1.9 ± 1.0 | – | 0.8 ± 0.5 | 0.2 ± 0.1[c] | |
| Sulfate (ferric sulfates) | 0.94 ± 0.08 | – | 0.7 ± 0.1 | – | |
| Sulfate ($H_2SO_4$ & ferric sulfates) | 2.89 ± 0.30 | – | 2.2 ± 0.3 | >2[d] | |
| Total $H_2O$ (hydrates and solution) | 29.49 ± 3.05 | – | 4.1 ± 0.7 | – | |
| (C) Mole Ratios | | | | | |
| | Fe : $SO_4$ (ferric sulfates) | $H_2O$ : ($H_3O$) jarosite | $H_2SO_4$ : ($H_3O$) jarosite | $H_2O$ : $H_2SO_4$ | – |
| Mole ratio | 2.0 ± 1.1 | 39 ± 22 | 3.0 ± 1.7 | 14 ± 3 | – |

[a] *e.g.*, Porshnev *et al.* (1987)
[b] Knollenberg and Hunten (1980)
[c] Andreichikov et al. (1987)
[d] Surkov et al. (1986)

~51–0 km).

$$Fe_2(SO_4)_3 + 3CO \rightarrow Fe_2O_3 + 3SO_2 + 3CO_2 \quad (17)$$

$$Fe_2(SO_4)_3 + 1.5CO + 0.5OCS \rightarrow Fe_2O_3 + 2CO_2 + 3.5SO_2 \quad (18)$$

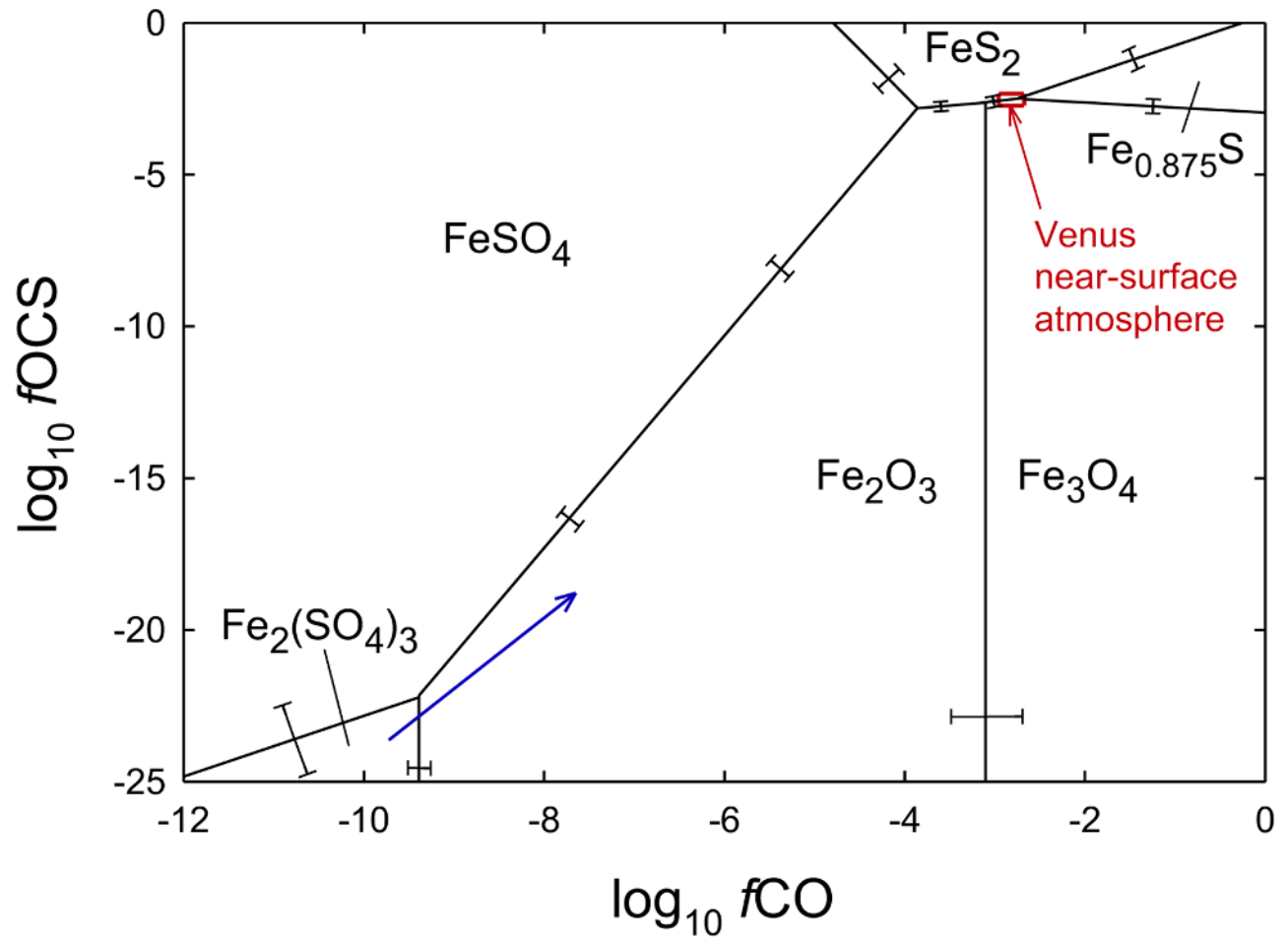

**Fig. 6.** Stability fields for solid phases in the Fe-O-S system at the surface of Venus (735 K, 92.1 bar, 6052 km) and volume mixing ratios of 0.965 $CO_2$ and 150 ppm $SO_2$. Mikasaite stability field is far from Venus' surface conditions. The red box corresponds to volume mixing ratios near the surface based on measurements (12–23 ppm CO) and models (20–50 ppm OCS). The blue arrow illustrates a potential mikasaite to hematite conversion via interactions with CO and OCS (Reactions (17) and (18)) in the lower atmosphere. Error bars depict uncertainties due to the thermodynamic data.

### 3.6. *Surface sourcing with gaseous iron halides*

We assessed the possibility that cloud iron could be delivered from the surface *via* iron halides that sublimate and rise to the clouds. Calculations of the gas-solid equilibria at surface conditions demonstrated that the ferric oxides are much more stable than the solid iron halides. Stability diagrams for the surface conditions (Fig. 7) yielded very low mixing ratios for $FeCl_{2(g)}$ (~$10^{-12}$, ~1 ppt), $FeCl_{3(g)}$ (~$10^{-16}$, ~0.1 ppq), $Fe_2Cl_4$, $Fe_2Cl_6$, FeCl, $Fe(CO)_5$, $FeF_3$, and $FeF_2$. These assessments considered (**a**) the low atmospheric mixing ratios for HCl (~0.4 ppm) and HF (~5 ppb), which are in insufficient in promoting the formation of substantial solid iron halides *via* the alteration of Fe-bearing phases (Zolotov, 2018; Zolotov, 2021), (**b**) the constrained equilibrium speciation of the C-O-N-S-H-Cl-F-Fe gas system, and (**c**) Fe/C mole ratios of $10^{-13}$ to $10^{-7}$. Calculations indicate that ~0.7 ppt $FeCl_{2(g)}$ and ~0.2 ppq $FeCl_{3(g)}$ would condense above ~29 and 58 km, respectively. Hence, only trace amounts of $FeCl_2$ and/or $FeCl_3$ gases and/or grains would be delivered to the clouds.

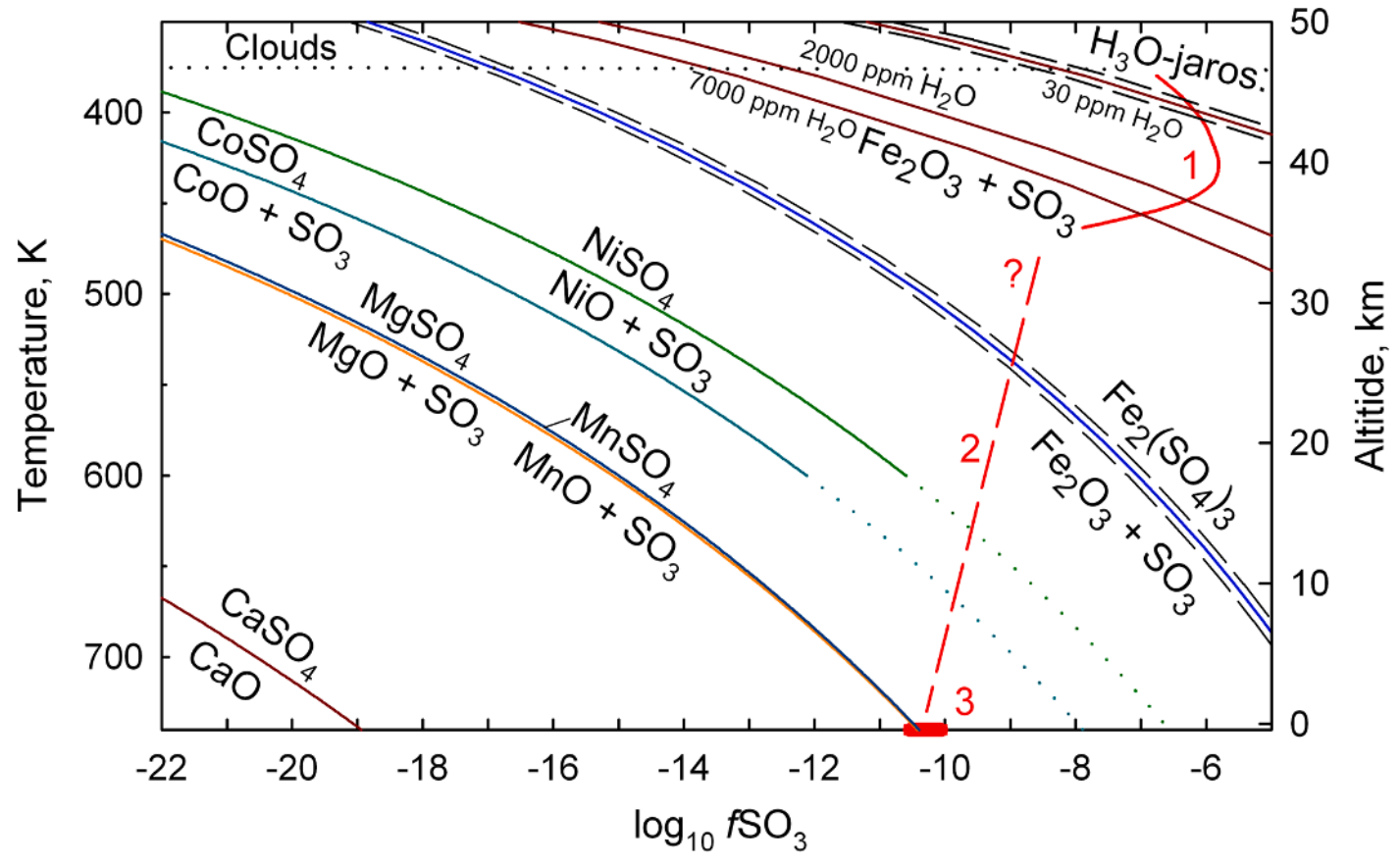

**Fig. 5.** Thermal stability of mikasaite, hydronium jarosite, and other anhydrous metal sulfates under the conditions of Venus' lower atmosphere. The respective compounds are stable above the solid curved lines and unstable below the lines. Uncertainties are represented by the associated dashed lines. The mikasaite-hematite phase boundary (blue curve), labeled as $Fe_2(SO_4)_3$ and $Fe_2O_3 + SO_3$, corresponds to Reaction (5). The equilibrium conditions for hydronium jarosite (magenta curves), labeled as $H_3O$-jaros., correspond to Reaction (8) in the presence of 30 ppm $H_2O$ (atmospheric vapor) and 2000 and 7000 ppm $H_2O$ (released $H_2O$ within the LNMS inlets). Other curves depict the metal sulfate-oxide-$SO_3$ equilibria for $NiSO_4$ (green), $CoSO_4$ (cyan), $MgSO_4$ (black), $MnSO_4$ (green), and $CaSO_4$ (purple). The non-dashed and dashed red lines show the $SO_3$ altitude profile, where the numbers in red font represent: (1) the $SO_3$ partial pressure at ~48–34 km (~0.2 ppm at ~40 km) based on the thermochemical kinetic model of Krasnopolsky (2013); (2) a hypothetical $SO_3$ altitude profile (demarked by the question mark) connecting the modeled values from Krasnopolsky (2013) at ~34 km (~20 ppb $SO_3$) to the surface, and (3) the near surface $SO_3$ partial pressure (~0.3 ppt $SO_3$) shown by the thick red line, based on gas-phase chemical equilibrium models (740 K, 95.6 bars, 6051.4 km) (Zolotov, 2026).

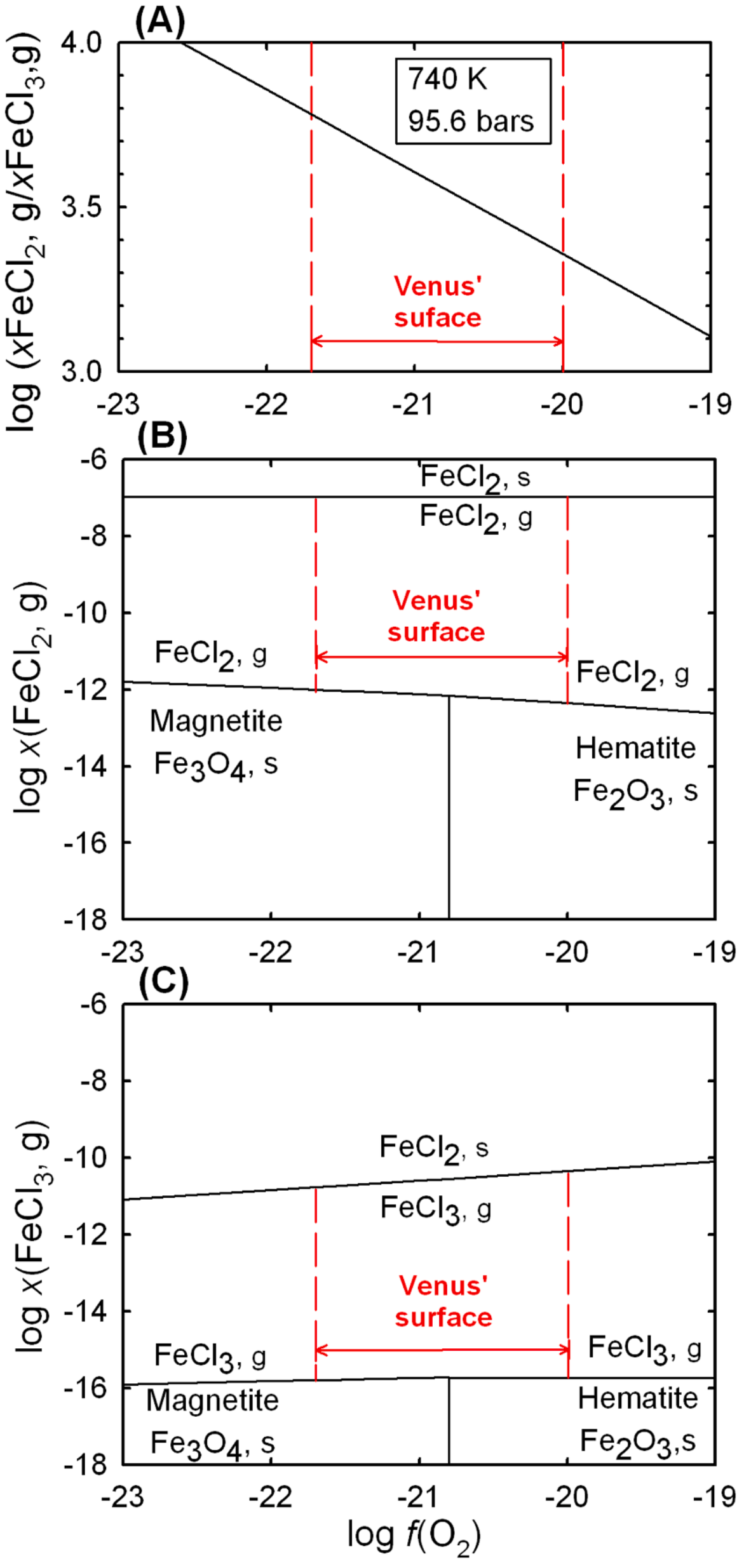

**Figure 7.** Volume mixing ratios ($x$) of iron chloride gases in chemical equilibria with ferric oxides at the conditions of Venus' modal radius surface (740 K, 95.6 bars, 6051.4 km). The $O_2$ fugacity ($fO_2$) in the near-surface atmosphere (red dashed lines) is within the uncertainty of the hematite-magnetite phase boundary of ~0.8 log $fO_2$ (Fegley *et al.*, 1997b). (A) The plot demonstrates the dominance of $FeCl_2$ over $FeCl_3$ in the gas phase, based on the $FeCl_2$-$FeCl_3$ equilibrium. (B) The plot shows the range of $FeCl_2$ mixing ratio limited by equilibria with iron oxides and condensation to solid $FeCl_2$. (E) The plot shows the range of $FeCl_3$ mixing ratio limited by equilibria with iron oxides and reduction to solid $FeCl_2$. Modified from Zolotov (2021).

## 4. Discussion

### 4.1. Iron in the cloud aerosols

The LNMS data support the presence of ferric sulfates in the aerosols captured at ~51–48 km within Venus' middle/lower clouds. Interpretation of the LNMS descent data as an evolved gas analysis suggests that the cloud aerosols thermally decomposed in the inlet after capture and released differing compounds into the mass analyzer along the atmospheric temperature gradient. At ≤23 km (inlet temperatures of ~560–740 K), the thermal decomposition products included $Fe_2O_3$, $SO_2$, $SO_3$, $O_2$, $H_2SO_4$, and $H_2O$. Several iron signals were suggested by the LNMS data, including $^{54}Fe^+$, $^{56}FeO^+$, $^{54}FeO^+$, $^{56}Fe_2O^+$, $^{56}Fe_2O_2^+$, and potentially $Fe_2O_3^+$, which were assigned as the electron ionization products of $Fe_2O_3$ (Reactions (10) and (11)).

The combined LNMS thermal evolution profiles (at ≤23 km; Fig. 4) suggest the decomposition of hydronium-containing ferric sulfates, possibly hydronium jarosite. The profiles for $Fe_2O_3$ and $SO_2$ generally track (*e.g.*, have similar $T_D$ values) within the uncertainties, supporting the decomposition of ferric sulfates. The profiles for $H_2SO_4$ and $SO_2$ generally track, suggesting that hydronium serves as a hydrogen source during the decomposition process. The dissimilar $H_2SO_4$ and $SO_3$ profiles (**Appendix Fig. A.1**) suggest different mechanisms of sulfate decomposition, possibly *via* contributions from hydronium. The broad $H_2O$ release step overlaps with the temperatures where $H_2O$ releases from hydronium jarosite (~523–673 K) (Spratt *et al.*, 2014).

The LNMS iron to sulfate mole ratio (2.0 ± 1.1) was obtained from the mass signals for $Fe_2O_3$ and $SO_2$ (at ≤23 km). Barring the presence of other iron species, the mole ratio suggests a formula of $(H_3O)Fe_3(SO_4)_{1.5}(OH)_7$. This is within the uncertainty limits to the iron to sulfate mole ratio (1.5) for hydronium jarosite ($(H_3O)Fe_3(SO_4)_2(OH)_6$). The LNMS mole ratio is outside of the uncertainty limits to volaschioite (4; $Fe_4(SO_4)O_2(OH)_6(H_2O)_2$), copiapite (0.83; $Fe^{2+}Fe^{3+}_4(SO_4)_6(OH)_2{\cdot}20H_2O$), mikasaite (0.67; $Fe_2(SO_4)_3$), and rhomboclase (0.5; $(H_3O)Fe(SO_4)_2$). Hence, the aerosol-derived iron-bearing species is referred to herein as hydronium jarosite.

The assignment of hydronium jarosite (or similar species) aligns with experimental observations. In Šolc *et al.* (1988) and Spratt *et al.* (2014), the thermal decomposition of hydronium jarosite was suggested to include stepwise losses of $Fe_2O_3$ (or Fe(III)O–bearing species). In Spratt *et al.* (2014), mass spectral studies showed the release of $SO_2$ in ~4 steps. In Frost *et al.* (2006) and Xu *et al.* (2010), $H_2SO_4$ was suggested as a thermal decomposition product of hydronium-bearing ferric sulfates (*e.g.*, hydronium jarosite and rhomboclase). The LNMS trends are consistent with these observations, including the release steps for $Fe_2O_3$ and $SO_2$, and the increases in $H_2SO_4$ vapor at the higher inlet temperatures (≥600 K).

However, the thermal decomposition in the LNMS inlet occurred at lower temperatures (~560–735 K) relative to the experimental studies (~520–920 K) (*e.g.*, Spratt *et al.* (2014)). Contributing factors likely included (**a**) the small sizes and low crystallinity of the sulfate grains that formed within the inlets, (**b**) low pressures in the mass analyzer (≤$10^{-8}$ bar), and (**c**) surface chemistry within the inlet. As originally described in Hoffman *et al.* (1980), the passage of $SO_3$ through the LNMS inlet yielded $SO_2$, solution-phase $H_2SO_4$ (85 wt%) did not pass through the inlet, and $H_2SO_4$ trapped in the inlet decomposed and released $SO_2$ under Venus-like conditions (~25–50 km).

Taken together, these results suggest that the aerosols from ~48–51 km contain ferric sulfates and that hydronium jarosite formed within the LNMS inlet after aerosol capture and thermal losses of sulfuric acid and water. This formation sequence is generally consistent with the synthesis of hydronium jarosite in Spratt *et al.* (2014), which involved heating an aqueous solution of ferric sulfate (~0.5 M) at 423 K and 4 bar. Alternatively, iron and sulfate from differing compounds or aerosols may have combined after capture. Regardless, the native speciation of ferric sulfate in the aerosols remains unconstrained and could include solution-phase ions, solution-phase neutral complexes, and/or suspended particles.

### 4.2. Mineral stability assessments

Thermochemical calculations support the assignment of ferric sulfates as a thermally unstable iron-bearing aerosol constituent, and hydronium jarosite as an aerosol-derived product. Under the sub-cloud conditions, ferric sulfates are the least stable among several tested metal sulfates (Fig. 5). Nickel and cobalt sulfates are more thermally stable than mikasaite and hydronium jarosite. Calcium, magnesium, sodium, and manganese sulfates would not appreciably decompose. Stability calculations were also conducted using the conditions within the LNMS inlet, which considered water vapor abundances in the inlet as the captured aerosols dehydrated across the descent. These calculations suggest that hydronium jarosite or mikasaite would be stable up ~36 and 25 km, respectively. Below these altitudes, the decomposition would yield $Fe_2O_3$ and $SO_3$ (Fig. 5). These models align with the LNMS descent data, which suggest that the thermal decomposition of the ferric sulfates began at ~23 km.

### 4.3. Aerosol composition

The LNMS mass loading values and relative abundances for the aerosol composition are generally consistent with other Venus measurements (Table 1). For bulk iron, the LNMS value (0.8 ± 0.5 mg $m^{-3}$) is higher than the tentative measurements from Venera 12 and Vega 2 (0.2 ± 0.1 mg $m^{-3}$) (Petryanov *et al.*, 1981; Andreichikov *et al.*, 1987), though potentially comparable due to the errors. For the total sulfate (from ferric sulfates and sulfuric acid), the LNMS value (2.2 ± 0.3 mg $m^{-3}$) is consistent with the Vega 1 pyrolysis mass spectrometry data (>2 mg $m^{-3}$) (Surkov *et al.*, 1986). These comparisons suggest that the PVLP, Venera 12, and Vega 1 and 2 sampled comparable aerosols. The total LNMS aerosol mass loading (7.4 ± 1.3 mg $m^{-3}$) aligns with the maximum mass loading measured at ~50–51 km (~12 mg $m^{-3}$) by the Pioneer Venus Large Probe Cloud Particle Size Spectrometer (Knollenberg and Hunten, 1980). The remaining mass in the aerosols may represent uncharacterized constituents.

Relative abundances of hydronium jarosite (30 ± 17 wt%), sulfuric acid (20 ± 5 wt%), and water (50 ± 10 wt%) in the captured aerosols were obtained using the thermal release profiles for $Fe_2O_3$, $SO_2$, and $H_2O$. By mass, sulfuric acid is the minor aerosol component. The aerosol $H_2O$ (47–51 km) represents ~0.3% of the total atmospheric $H_2O$ vapor (~30 ppmv, 0–70 km) (see **Appendix B**). When considering only the liquid-phase components, per Mogul *et al.* (2025), relative abundances are $H_2SO_4$ (86 ± 17 wt%) and $H_2O$ (14 ± 4 wt%), consistent with other Venus measurements, as reviewed in Dai *et al.* (2025), Mogul *et al.* (2025), and Zolotov (2026).

The presence of other aerosol compounds is suggested by the LNMS data. The $H_2O$ and $O_2$ profiles do not necessarily track with $SO_2$, suggesting the presence of additional decomposing aerosol species. The substantial $H_2O$ that released at ~548–735 K yielded an apparent mole ratio of water to hydronium jarosite (39 ± 22) that suggests the presence of other hygroscopic and/or water-releasing compounds in the aerosols. Plausible candidates that thermally release water at ~548–735 K, as described in Mogul *et al.* (2025), include kieserite ($MgSO_4 \cdot H_2O$), kaolinite ($Al_2Si_2O_5(OH)_4$), perlite and/or opal (*e.g.*, amorphous $SiO_2 \cdot nH_2O$), and/or insoluble organic matter (IOM), which releases $H_2O$ *via* the decomposition of hydroxide and/or carbonyl groups. Qualitative support for such species includes the LNMS signals potentially consistent with $MgSO_4^+$, $SiO_2^+$, $Si_2O^+$, $SiO^+$, and $Al_2O_3^+$ (Zolotov *et al.*, 2023; Mogul *et al.*, 2025). The abundances of these compounds are currently unknown and not included in the mass percent values.

### 4.4. Cosmic delivery of iron to the clouds and surface

We suggest that the cloud iron and other aerosol-phase materials (barring sulfuric acid) are delivered into Venus' atmosphere *via* the influx of cosmic dust, which is mostly cometary in origin (Carrillo-Sánchez *et al.*, 2020). Expected alteration products of cometary dust include ferric sulfates, nickel and cobalt sulfates, kieserite, kaolinite, perlite/opal, and IOM – when considering ablation above the clouds and reactions with the sulfuric acid aerosols (this work, Mogul *et al.* (2025), and Zolotov *et al.* (2023)). This assessment aligns with the microphysics model in Karyu *et al.* (2026) regarding cosmic dust infall into Venus' atmosphere, the rates of cosmic dust infall in Carrillo-Sánchez *et al.* (2020), and the chemical model for the formation of ferric chloride from ablated iron at the cloud tops in Egan *et al.* (2026).

Our calculations suggest a geologically short residence time for the cloud aerosol iron (2.4 ± 1.5 Myr), when assuming a steady state of cloud influx and aerosol fallout from the clouds. Ultimately, aerosol fallout could lead to substantial recycling of sulfuric acid vapor back to the clouds (Dai *et al.*, 2022) and settling of the non-volatile components. Near the surface (≤25 km), the ferric sulfates would thermally decompose, yielding hematite. The released $SO_2$ and $H_2O$ would be negligible against the respective atmospheric gas abundances (~150 and 30 ppmv). As a bulk calculation, the thermal decomposition of the total aerosols from 47–51 km would yield trace amounts of $SO_2$ (~$10^{12}$ kg) and $H_2O$ (~$10^{13}$ kg) compared to the atmospheric masses of $SO_2$ (~$10^{17}$ kg) and $H_2O$ (~$10^{15}$ kg) at ≤47 km, respectively (**Appendix B**). In contrast, atmospheric settling of the iron over ~700 Myr, or since the offset of global volcanism, could yield a surface hematite layer with a 0.43 ± 0.25 mm thickness. The near-infrared spectra of Venus' surface at ~1 μm (*e.g.*, Pieters *et al.* (1986)) could thus be consistent with the presence of hematite formed from cosmic sources (this work and Zolotov (2021)) and through the chemical weathering of native ferrous minerals and glasses (*e.g.*, Fegley *et al.* (1997a) and Zolotov (2018)).

### 4.5. Potential for surface sourcing of cloud iron

We assessed the potential for particles and gases to deliver iron to the clouds. Basalts, the dominant venusian surface rocks, contain 6.5 ± 0.8 wt% Fe based on data from Venera and Vega landers (*e.g.*, Head *et al.* (1992) and Filiberto *et al.* (2025)). Fragmented basaltic surface material was observed at the four Venera landing sites and inferred from Magellan radar images (*e.g.*, Basilevsky *et al.* (1985), Basilevsky *et al.* (2004), and Carter *et al.* (2023)). The presence of basaltic dust in the near-surface haze layer was inferred from Venera 13 data (Kulkarni *et al.*, 2025). Models for the near-surface winds suggest lateral transport of mineral grains (Lefèvre *et al.*, 2025). However, the delivery of basaltic dust to the clouds remains to be assessed, including the effects of topography-driven atmospheric waves and Hadley cell-type upwellings.

On Earth, volcanic ash supplies iron to the atmosphere (*e.g.*, Hull *et al.* (2025)). However, at Venus, volcanic morphologies are inconsistent with plume deposition (Head et al., 1992). Large volcanic plumes are not expected at the near-surface atmospheric pressures (~90-100 bar), which suppress magma degassing, pyroclastic activity, and plume formation (*e.g.*, Airey *et al.* (2015) and Gaillard *et al.* (2021)). While iron-bearing gases are detected and modeled in volcanic fumaroles on Earth (*e.g.*, Mandon *et al.* (2025)) and assessed in basaltic volcanic gases on the Moon (Renggli *et al.*, 2017) and Venus (Zolotov, 2021), iron is among the least abundant metals in Earth's basaltic volcanic gases, not relevant to arc volcanism (Edmonds *et al.*, 2022). Moreover, the age of Venus' surface (~0.7 Ga) and ambiguity regarding current volcanic activity (Filiberto *et al.*, 2025) do not imply the emission of Fe-bearing volcanic gases in the present epoch.

Krasnopolsky (2017) suggested that cloud iron could be supplied by gaseous ferric chloride(s) formed *via* the sublimation of surface materials. However, the model in Woitke *et al.* (2025) indicated only trace abundances of gaseous iron chloride below the clouds. Our work shows that the solid-phase iron halides are less stable than the iron oxides and sulfides (hematite, magnetite, and pyrite), which do not sublimate under Venus conditions and are the major secondary Fe-bearing surface minerals (Zolotov, 2018). Our equilibrium concentrations of the iron chloride gases are negligible (~ 1 ppt) in the presence of these major minerals (Fig. 7). The presence of hematite is suggested by near-infrared reflectance measurements at the Venera landing sites (Pieters et al., 1986). Additionally, the cloud delivery of iron chloride gases could be impeded by conversion to to $Fe_2Cl_{6(g)}$ at the lower altitudes, which could prevent condensation (Krasnopolsky, 2017), or to pyrite in the highlands and above (Zolotov, 2018; Woitke *et al.*, 2025). Thus, surface sourcing *via* Fe-bearing gases is unlikely, whereas sourcing *via* dust particles remains unconstrained.

## 5. Conclusion

The LNMS data suggest that the aerosols in Venus' middle/lower clouds (~48–51 km) contain abundant ferric sulfates. This revises the perception that the aerosol composition is dominated by sulfuric acid. The LNMS data were treated as an evolved gas analysis of the thermal decomposition of aerosols, which were inadvertently captured in the intake inlets and subjected to the atmospheric temperature gradient of the descent. The thermal decomposition products that released into the LNMS (at ≤23 km) are consistent with hydronium jarosite (or similar compound), which likely formed within the LNMS inlet after aerosol capture and losses of sulfuric acid and water. Hence, speciation of the ferric sulfates in the native cloud aerosols remains unconstrained (*e.g.*, solution and/or particle phases). We propose that the aerosol-phase iron is sourced by cosmic dust infall, chemically altered to ferric sulfates *via* reactions with sulfuric acid, and thermally decomposed to ferric oxides after fallout from the clouds. Therefore, our results lend support to an iron altitude profile that includes ferric chloride and sulfates as UV absorbers in the cloud tops and aerosols, respectively, ferric sulfates as components of the sub-cloud hazes, and ferric oxides as components of the lower atmospheric hazes and surface materials.

## Data and Materials Availability

The LNMS data are available for download at the NASA Planetary Data System (Belkhodja *et al.*, 2025). The laboratory notebooks of J. H. Hoffman (Principal Investigator in the original LNMS project) from 1976-1979 are available for download as a collection on Zenodo (Hoffman and Mogul, 2025). The Hoffman laboratory notebooks were obtained from the archives of the William B. Hanson Center for Space Sciences at the University of Texas, Dallas. The PVLP project report used in the study (Final Report, 1978) can be reviewed upon request at the NASA Ames Research Center Archives as part of the Pioneer Project Collection (https://www.nasa.gov/archives/arc-archives/#CollectionsOvw-arc).

## CRediT authorship contribution statement

**Rakesh Mogul:** Writing – review & editing, Writing – original draft, Visualization, Project administration, Methodology, Investigation, Funding acquisition, Formal analysis, Data curation, Conceptualization. **Mikhail Yu Zolotov:** Writing – review & editing, Visualization, Methodology, Investigation, Formal analysis. **Michael J. Way:** Writing – review & editing, Investigation, Data curation. **Sanjay S. Limaye:** Writing – review & editing.

## Declaration of competing interest

The authors declare they have no known competing financial or personal relationships that could have appeared to influence the work reported in this paper.

## Acknowledgements

RM, MYZ, MJW, and SSL acknowledge support from the NASA Solar Systems Working program (award 80NSSC24K0929). MYZ was also supported by the NASA Discovery program (award 80NSSC22M0187). MJW was also supported through the NASA Astrobiology Program, NExSS, NASA Habitable Worlds Program, and the GSFC Sellers Exoplanet Environments Collaboration. RM acknowledges administrative support from Blue Marble Space.

## Appendix A

### *Extracting the LNMS chlorine isotope composition*

The Venus $^{35}Cl/^{37}Cl$ ratio was extracted from the LNMS cloud data at 58.3–50.3 km. The counts for $^{35}Cl^+$ and $^{37}Cl^+$ were acquired by refining the multi-step deconvolution procedure from Mogul *et al.* (2021) and using the peak-fitting procedures from Mogul *et al.* (2025). These steps are summarized in the following numbered list.

(1) Peak fits at 34 u deconvoluted the counts for isobaric $H_2S^+$ and $PH_3^+$ (this mixture of isobars yielded the best fits at 34 u).
(2) Peak fits at 35 u deconvoluted the counts for isobaric $^{35}Cl^+$, $HDS^+$, and $PH_2D^+$; where $HDS^+$ and $PH_2D^+$ were constrained using $H_2S^+$ and $PH_3^+$ and a D/H ratio range of 0.005–0.2.
(3) The parent species for $^{35}Cl^+$ was defined as HCl, the likely major Cl-bearing species in the clouds.
(4) Counts for $HCl^+$ were obtained by multiplying $^{35}Cl^+$ by the $HCl^+/Cl^+$ ratio from the NIST database (~5.56) (Wallace, 2020).
(5) Counts for $DCl^+$ were obtained by multiplying $HCl^+$ by the DCl/HCl ratio from 74 km (0.0303) from Krasnopolsky *et al.* (2013).
(6) Counts for $^{37}Cl^+$ were obtained by subtracting $DCl^+$ from the total counts at 37 u.
(7) The deconvoluted counts for $^{35}Cl^+$ and $^{37}Cl^+$ yielded the $^{35}Cl/^{37}Cl$ ratio.
(8) The $^{35}Cl/^{37}Cl$ ratio and uncertainty were assigned as the average and standard deviation from the 4 spectra obtained at 50.3–58.3 km.

## Appendix B

### *Estimating the masses of iron, sulfate, and water in the cloud aerosol layer*

The major iron aerosol layer was defined as 47 to 51 km. This correlated to where the capture of clog-inducing aerosols by the LNMS began during descent (~51 km) and the cloud deck ends (~47 km). Across this range, we estimated the total masses of iron ($m_{Fe}$), sulfate from ferric sulfates, sulfate from $H_2SO_4$ and ferric sulfates, $H_2O$ from the solution and hydrate phases, and $SO_2$ produced from the thermal decomposition. We assumed a constant mass percent (wt%) of the aerosol constituents against the total aerosol mass loading. This permitted a means to account for the large range in aerosol mass loadings (~0.01 to 80 mg $m^{-3}$) that were observed between ~47 and 51 km (Appendix Fig. B1) by the Pioneer Venus Large Probe Cloud Particle Size Spectrometer (LCPS) (Knollenberg and Hunten, 1980).

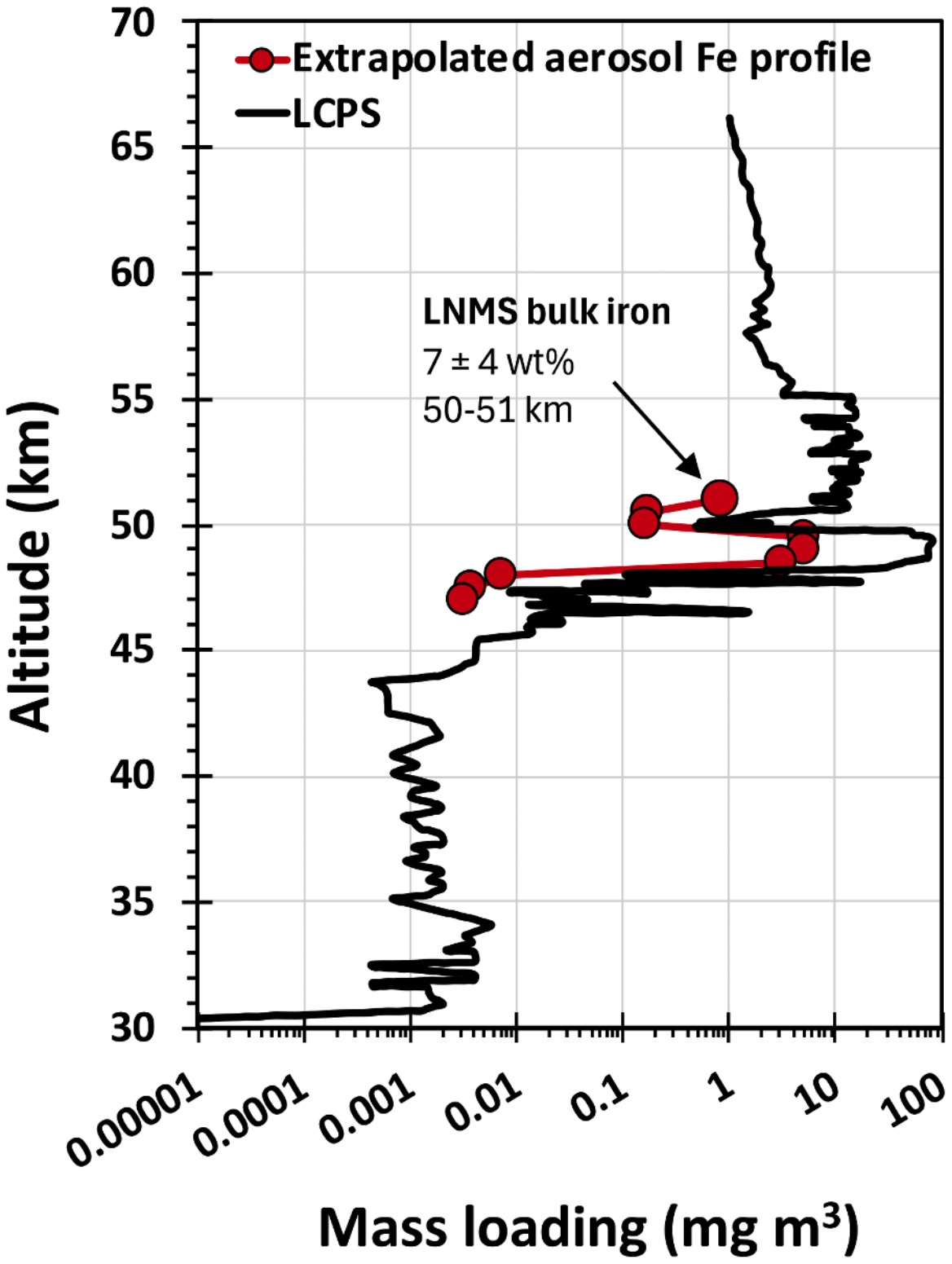

**Fig. B.1.** Aerosol mass loading from the LCPS (black line) and aerosol iron mass loading at 47-51 km extrapolated from the LNMS data (red line; markers represent the 0.5 km increments used in the calculations.

The aerosol mass percent values were calculated against a reference aerosol mass loading, which was assigned as the maximum value measured between 50 and 51 km by the LCPS (12 ± 1 mg $m^{-3}$) (Knollenberg and Hunten, 1980). Compared to this value, the LNMS mass loading for iron (0.8 ± 0.5 mg $m^{-3}$) translated to an aerosol mass percent of 7 ± 4 wt%. The LNMS mass loading for sulfate from ferric sulfates (0.7 ± 0.1 mg $m^{-3}$) translated to an aerosol mass percent of 6 ± 1 wt%. The LNMS mass loading for total sulfate from $H_2SO_4$ and ferric sulfates (2.2 ± 0.3 mg $m^{-3}$) translated to an aerosol mass percent of 18 ± 3 wt%. The LNMS mass loading for $H_2O$ from the solution and hydrate phases (4.1 ± 0.7 mg $m^{-3}$) translated to an aerosol mass percent of 35 ± 7 wt%. The LNMS values amounted to a total mass loading of 7.4 ± 1.3 mg $m^{-3}$. Compared the maximum LCPS mass loading (12 ± 1 mg $m^{-3}$) at 50–51 km, the remaining ~34 wt% of the aerosol mass was assumed to represent uncharacterized volatile and/or non-volatile compounds.

The resulting iron mass loading profile from 47–51 km, when assuming a constant mass percent, is shown in Appendix Fig. B1 (in 0.5 km increments). Multiplication of these mass loading values by the shell volumes (modal planetary radius of 6051.4 km), followed by summation of the shell masses, yielded the estimated total iron mass. The sulfate and water masses were similarly obtained. The total masses at 47–51 km (~$1.9\times10^{18}$ $m^3$) were obtained for bulk iron ($3.5\times10^{12} \pm 1.0\times10^{12}$ kg), sulfate from ferric sulfates ($2.9\times10^{12} \pm 0.3\times10^{12}$ kg), sulfate from $H_2SO_4$ and ferric sulfates ($9.0\times10^{12} \pm 0.7\times10^{12}$ kg), and $H_2O$ from the solution and hydrate phases ($1.7\times10^{13} \pm 0.2\times10^{13}$ kg). Total masses of sulfate were also converted to the masses of $SO_2$ that would release from ferric sulfates ($1.9\times10^{12} \pm 0.2\times10^{12}$ kg) and $H_2SO_4$ and ferric sulfates ($6.0\times10^{12} \pm 0.5\times10^{12}$ kg) after thermal decomposition.

Atmospheric masses of 150 ppmv $SO_2$ and 30 ppmv $H_2O$ were also obtained. Mixing ratios were converted to mass loading (in 1 km increments) using the pressure and temperature profile of the atmosphere (Seiff *et al.*, 1985). Mass loading values were converted to shell mass using the shell volume. Shell mass as a function of altitude was fitted with a fourth-order polynomial. Integration using defined altitude ranges yielded the total masses. At ≤47 km, the atmospheric masses were $SO_2$ (~$10^{17}$ kg) and $H_2O$ (~$10^{15}$ kg).

Comparison of these values suggest that the clog-inducing aerosols represent trace fractions of the atmosphere. The total aerosol $H_2O$ (predominantly from the hydrate phase) constitutes ~0.3% of the total atmospheric $H_2O$ vapor (30 ppmv) from 1–70 km. The thermal evaporation and decomposition of the total aerosols at 47–51 km would yield $H_2O$ abundances (~$10^{13}$) that are negligible compared to atmospheric $H_2O$ abundances at ≤47 km (~$10^{15}$ kg). The thermal decomposition of the total $H_2SO_4$ and ferric sulfates at 47–51 km would yield $SO_2$ abundances (~$10^{12}$ kg) that are negligible compared to atmospheric $SO_2$ abundances at ≤47 km (~$10^{17}$ kg).

## Appendix C

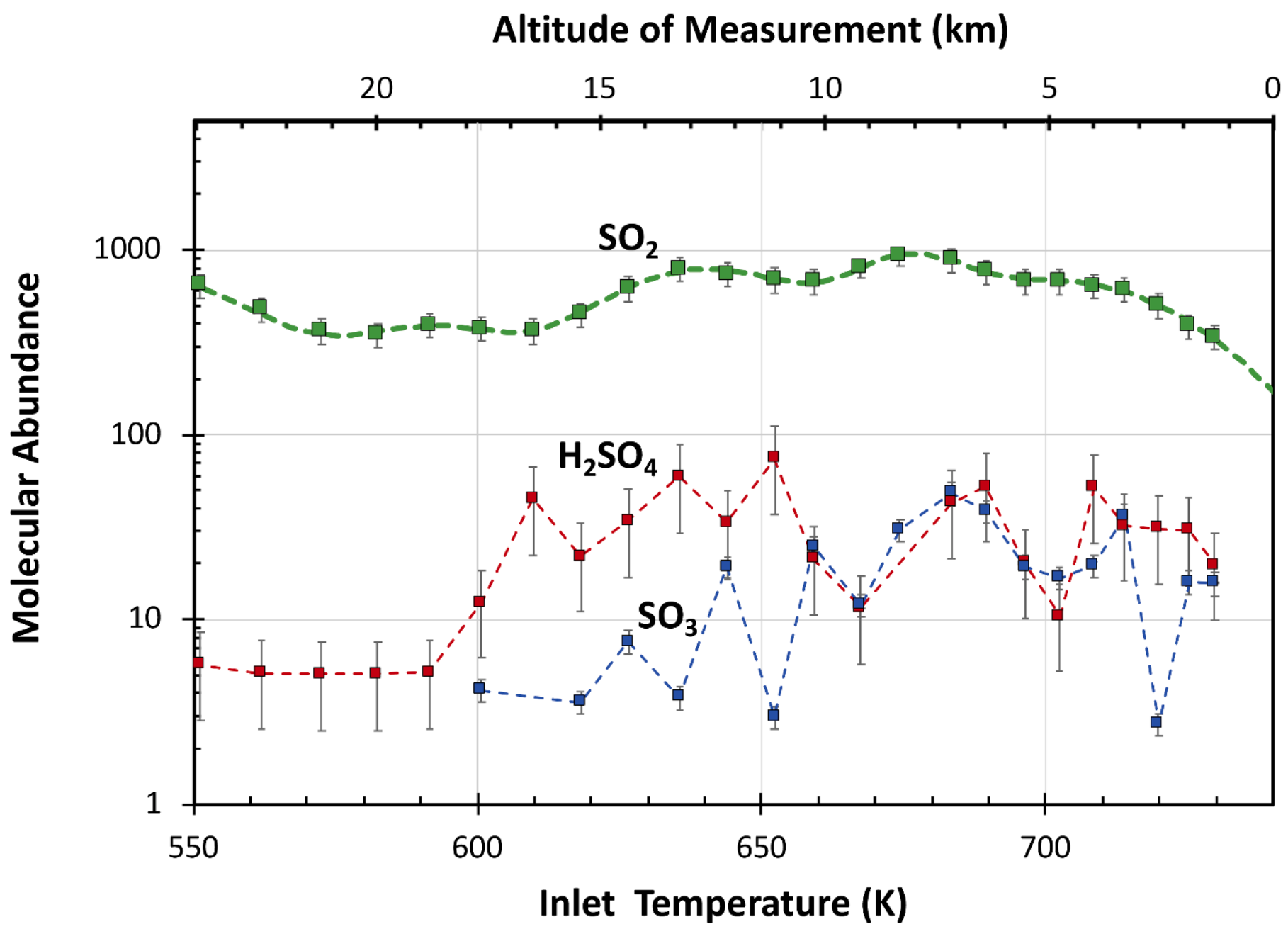

**Fig. C.1.** Overlay of the $SO_2$, $SO_3$, and $H_2SO_4$ thermal evolution profiles associated with hydronium jarosite decomposition. Release of $H_2SO_4$ occurred at ~600–670, 670–700, and 700–740 K. Release of $SO_3$ occurred at ~626, 644, 650–700, 700–720, and 720–740 K.

**Table C.1**
The LNMS aerosol composition, when including mikasaite, sulfuric acid, and water. Aerosols were captured at ~51-48 km. Compositional parameters include the number of released molecules (*N*), weight percentage (wt%), and mass loading (mg $m^{-3}$). Values for $H_2SO_4$ are from Mogul *et al.* (2025).

| LNMS Aerosol Composition | | | |
|---|---|---|---|
| Mikasaite, Sulfuric Acid, & Water | | | |
| Compounds | N ($\times 10^5$) | wt% | mg $m^{-3}$ |
| aerosol $H_2O$ | 29.49 ± 3.05 | 48 ± 11 | 4.1 ± 0.7 |
| $Fe_2(SO_4)_3$ | 0.96 ± 0.41 | 35 ± 20 | 3.0 ± 1.6 |
| $H_2SO_4$ | 1.95 ± 0.28 | 17 ± 4 | 1.5 ± 0.2 |
| Total mass loading | | | 8.6 ± 1.7 |

## References

Airey, M.W., Mather, T., Pyle, D., Glaze, L., Ghail, R., Wilson, C., 2015. Explosive volcanic activity on Venus: the roles of volatile contribution, degassing, and external environment. Planet. Space Sci. 113, 33–48. https://doi.org/10.1016/j.pss.2015.01.009.

Andreichikov, B.M., Akhmetshin, I., Korchuganov, B., Mukhin, L., Ogorodnikov, B., 1987. X-ray radiometric analysis of Venus cloud aerosol by VEGA-1 and -2 automated interplanetary probe. Cosm. Res. 25, 721–736.

Basilevsky, A.T., Head, J.W., Abdrakhimov, A.M., 2004. Impact crater air fall deposits on the surface of Venus: areal distribution, estimated thickness, recognition in surface panoramas, and implications for provenance of sampled surface materials. J. Geophys. Res. Planets 109. https://doi.org/10.1029/2004JE002307.

Basilevsky, A.T., Kuzmin, R.O., Nikolaeva, O.V., Pronin, A.A., Ronca, L.B., Avduevksy, V. S., Uspensky, G.R., Cheremukhina, Z.P., Semendhenko, V.V., Ladygin, V.M., 1985. The surface of Venus as revealed by the Venera landings: Part II. GSA Bulletin 96, 137–144. https://doi.org/10.1130/0016-7606(1985)96<137:TSOVAR>2.0.CO;2.

Bastian, R., Weberling, R., Palilla, F., 1953. Spectrophotometric Determination of Ion as Ferric Sulfate Complex. Anal. Chem. 25, 284–288. https://doi.org/10.1021/ac60074a020.

Beard, B.L., Johnson, C.M., 2004. Fe isotope variations in the modern and ancient earth and other planetary bodies. Rev. Mineral. Geochem. 55, 319–357. https://doi.org/10.2138/gsrmg.55.1.319.

Belkhodja, I., Williams, D.R., Hoffman, J.H., Hodges, R.R., Seiff, A., 2025. Pioneer venus large probe neutral mass spectrometer gas sampling data (1-15 amu and 15-208 amu) bundle. NASA Planetary Data System. https://doi.org/10.17189/n23p-9r21.

Brand, W.A., Coplen, T.B., Vogl, J., Rosner, M., Prohaska, T., 2014. Assessment of international reference materials for isotope-ratio analysis (IUPAC Technical Report). Pure Appl. Chem. 86, 425–467. https://doi.org/10.1515/pac-2013-1023.

Broecker, W.S., Peng, T.-H., 1982. Tracers in the Sea. Eldigio Press, Palisades, NY.

Carrillo-Sánchez, J.D., Gómez-Martín, J.C., Bones, D.L., Nesvorný, D., Pokorný, P., Benna, M., Flynn, G.J., Plane, J.M., 2020. Cosmic dust fluxes in the atmospheres of Earth, Mars, and Venus. Icarus 335, 113395. https://doi.org/10.1016/j.icarus.2019.113395.

Carter, L.M., Gilmore, M.S., Ghail, R.C., Byrne, P.K., Smrekar, S.E., Ganey, T.M., Izenberg, N., 2023. Sedimentary Processes on Venus. Space Sci. Rev. 219, 85. https://doi.org/10.1007/s11214-023-01033-2.

Crisp, D., 1986. Radiative forcing of the Venus mesosphere: I. Solar fluxes and heating rates. Icarus 67, 484–514. http://www.sciencedirect.com/science/article/pii/0019103586901260.

Dai, L., Titov, D.V., Shao, W.D., Zhang, X., Cui, J., Fan, S., 2025. Venus cloud research: progress and perspectives. Space Sci. Rev. 221, 51. https://doi.org/10.1007/s11214-025-01176-4.

Dai, L., Zhang, X., Shao, W.D., Bierson, C.J., Cui, J., 2022. A Simple Condensation Model for the $H_2SO_4$-$H_2O$ Gas-Cloud System on Venus. J. Geophys. Res. Planets 127, e2021JE007060. https://doi.org/10.1029/2021JE007060.

Edmonds, M., Mason, E., Hogg, O., 2022. Volcanic outgassing of volatile trace metals. Ann. Rev. Earth Planet. Sci. 50, 79–98. https://www.annualreviews.org/content/journals/10.1146/annurev-earth-070921-062047.

Egan, J.V., Feng, W., James, A.D., Manners, J., Marsh, D.R., Lebonnois, S., Lefèvre, F., Stolzenbach, A., Plane, J.M., 2026. Ferric chloride: the Venusian unknown UV absorber? J. Geophys. Res. Planets 131, e2025JE009634. https://doi.org/10.1029/2025JE009634.

Egan, J.V., James, A.D., Plane, J.M.C., 2025. Laboratory measurements of ferric chloride ($FeCl_3$) under Venusian conditions. ACS Earth Space Chem. https://doi.org/10.1021/acsearthspacechem.5c00132.

Fegley Jr., B., Klingelhöfer, G., Lodders, K., Widemann, T., 1997a. Geochemistry of surface-atmosphere interactions on Venus. Venus II: Geology, Geophysics, Atmosphere, and Solar Wind Environment, p. 591.

Fegley Jr., B., Zolotov, M.Y., Lodders, K., 1997b. The oxidation state of the lower atmosphere and surface of Venus. Icarus 125, 416–439. https://doi.org/10.1006/icar.1996.5628.

Filiberto, J., Zolotov, M.Y., Kohler, E., D'Incecco, P., Gorinov, D.A., Bhiravarasu, S.S., Weller, M.B., Brossier, J.F., López, I., Mason, P.J., Edmond, J.A., Mari, N., Komatsu, G., Di Achille, G., Garvin, J.B., 2025. Assessing the evidence for active volcanism on Venus: current limitations and prospects for future investigations. Geochem 85, 126316. https://www.sciencedirect.com/science/article/pii/S0009281925000716.

Final Report, 1978. Large probe neutral mass spectrometer [Report]. NASA Ames Research Center History Archives. Collection Number AFS8100.15A.0000261.

Frankland, V.L., James, A.D., Carrillo-Sánchez, J.D., Nesvorný, D., Pokorný, P., Plane, J.M.C., 2017. CO oxidation and $O_2$ removal on meteoric material in Venus' atmosphere. Icarus 296, 150. https://doi.org/10.1016/j.icarus.2017.06.005.

Frost, R., Wills, R., Kloprogge, J., Martens, W., 2006. Thermal decomposition of hydronium jarosite $(H_3O)Fe_3(SO_4)_2(OH)_6$. J. Therm. Analysis Calorim. 83, 213–218. https://doi.org/10.1007/s10973-005-6908-0.

Gaillard, F., Bouhifd, M.A., Füri, E., Malavergne, V., Marrocchi, Y., Noack, L., Ortenzi, G., Roskosz, M., Vulpius, S., 2021. The diverse planetary ingassing/outgassing paths produced over billions of years of magmatic activity. Space Sci. Rev. 217, 22. https://doi.org/10.1007/s11214-021-00802-1.

Head, J.W., Crumpler, L.S., Aubele, J.C., Guest, J.E., Saunders, R.S., 1992. Venus volcanism: Classification of volcanic features and structures, associations, and global distribution from Magellan data. J. Geophys. Res. Planets 97, 13153–13197. https://doi.org/10.1029/92JE01273.

Hoffman, J.H., Hodges, R.R., Donahue, T.M., McElroy, M.B., 1980. Composition of the Venus lower atmosphere from the Pioneer Venus mass spectrometer. J. Geophys. Res. Space Phys. 85, 7882–7890. https://doi.org/10.1029/JA085iA13p07882.

Hoffman, J.H., Mogul, R., 2025. Laboratory notebooks for the pioneer venus large probe neutral mass spectrometer [collection]. Zenodo. https://doi.org/10.5281/zenodo.15160195.

Huber, S.E., Mauracher, A., Sukuba, I., Urban, J., Maihom, T., Probst, M., 2017. Electron impact ionisation cross sections of iron oxides. Eur. Phys. J. D 71, 1–8. https://doi.org/10.1140/epjd/e2017-80308-2.

Hull, T., D'Aronco, S., Crumeyrolle, S., Hanoune, B., Giammanco, S., La Spina, A., Salerno, G., Soldà, L., Badocco, D., Pastore, P., Sellitto, P., Giorio, C., 2025. Metal speciation of volcanic aerosols from Mt. Etna at varying aerosol water content and pH obtained by different thermodynamic models. Environ. Sci.: Atmospheres 5, 8–24. https://doi.org/10.1039/D4EA00108G.

Jiang, C.Z., Rimmer, P.B., Lozano, G.G., Tosca, N.J., Kufner, C.L., Sasselov, D.D., Thompson, S.J., 2024. Iron-sulfur chemistry can explain the ultraviolet absorber in the clouds of. Venus. Sci. Adv. 10, eadg8826. https://doi.org/10.1126/sciadv.adg8826.

Karyu, H., Kuroda, T., Määttänen, A., Mahieux, A., Viscardy, S., Terada, N., Robert, S., Vandaele, A.C., Crucifix, M., 2026. A cosmic origin of Venus' lower haze. Nat. Astron. https://doi.org/10.1038/s41550-026-02843-4.

Knollenberg, R., Hunten, D., 1980. The microphysics of the clouds of Venus: results of the Pioneer Venus particle size spectrometer experiment. J. Geophys. Res. Space Phys. 85, 8039–8058. https://doi.org/10.1029/JA085iA13p08039.

Krasnopolsky, V.A., 2013. $S_3$ and $S_4$ abundances and improved chemical kinetic model for the lower atmosphere of Venus. Icarus 225, 570. https://ui.adsabs.harvard.edu/abs/2013Icar..225..570K.

Krasnopolsky, V.A., 2017. On the iron chloride aerosol in the clouds of Venus. Icarus 286, 134–137. https://doi.org/10.1016/j.icarus.2016.10.003.

Krasnopolsky, V.A., Belyaev, D.A., Gordon, I.E., Li, G., Rothman, L.S., 2013. Observations of D/H ratios in $H_2O$, HCl, and HF on Venus and new DCl and DF line strengths. Icarus 224, 57–65. http://www.sciencedirect.com/science/article/pii/S0019103513000705.

Kuiper, G., Sill, G., 1969. Identification of the Venus cloud layers. Bull. Am. Astronom. Soc. 351.

Kulkarni, S.V., Irwin, P.G.J., Wilson, C.F., Ignatiev, N.I., 2025. A search for the near-surface particulate layer using Venera 13 *in situ* spectroscopic observations. J. Geophys. Res. Planets 130, e2024JE008728. https://doi.org/10.1029/2024JE008728.

Lefèvre, M., Lebonnois, S., Spiga, A., Forget, F., 2025. The effect of near-surface winds on surface temperature and dust transport on Venus. J. Geophys. Res. Planets 130, e2025JE009133. https://doi.org/10.1029/2025JE009133.

Li, M., Liu, S.-R., Armentrout, P., 2009. Collision-induced dissociation studies of $Fe_mO_n^+$: Bond energies in small iron oxide cluster cations, $Fe_mO_n^+$ (m= 1–3, n= 1–6). J. Chem. Phys. 131. https://doi.org/10.1063/1.3246840.

Magenheim, A.J., Spivack, A.J., Volpe, C., Ransom, B., 1994. Precise determination of stable chlorine isotopic ratios in low-concentration natural samples. Geochim. Cosmochim. Acta 58, 3117–3121. https://doi.org/10.1016/0016-7037(94)90183-X.

Majzlan, J., Navrotsky, A., Stevens, R., Donaldson, M., Woodfield, B.F., Boerio-Goates, J., 2005. Thermodynamics of monoclinic $Fe_2(SO_4)_3$. J. Chem. Thermodyn. 37, 802–809. https://doi.org/10.1016/j.jct.2004.11.021.

Majzlan, J., Stevens, R., Boerio-Goates, J., Woodfield, B.F., Navrotsky, A., Burns, P.C., Crawford, M.K., Amos, T.G., 2004. Thermodynamic properties, low-temperature heat-capacity anomalies, and single-crystal X-ray refinement of hydronium jarosite, $(H_3O)Fe_3(SO_4)_2(OH)_6$. Phys. Chem. Minerals 31, 518–531. https://doi.org/10.1007/s00269-004-0405-z.

Mandon, C.L., Kaasalainen, H., Calabrese, S., Shock, E.L., Prapaipong, P., Tassi, F., Gunnarsson, I., Gunnarsson-Robin, J., Stefánsson, A., 2025. Trace element transport by volcanic gases at Vulcano (Sicily, Italy) – speciation, deposition and fluxes. J. Volcanol. Geothermal Res. 458, 108235. https://www.sciencedirect.com/science/article/pii/S0377027324002282.

Markiewicz, W.J., Petrova, E., Shalygina, O., Almeida, M., Titov, D.V., Limaye, S.S., Ignatiev, N., Roatsch, T., Matz, K.D., 2014. Glory on Venus cloud tops and the unknown UV absorber. Icarus 234, 200–203. https://doi.org/10.1016/j.icarus.2014.01.030.

Mills, F.P., Esposito, L.W., Yung, Y.L., 2007. Atmospheric Composition, Chemistry, and Clouds, Exploring Venus as a Terrestrial Planet. Wiley, pp. 73–100. https://agupubs.onlinelibrary.wiley.com/doi/abs/10.1029/176GM06.

Mogul, R., Limaye, S.S., Way, M., Cordova, J.A., 2021. Venus' mass spectra show signs of disequilibria in the middle clouds. Geophys. Res. Lett. 48, e2020GL091327. https://doi.org/10.1029/2020GL091327.

Mogul, R., Limaye, S.S., Way, M.J., 2023. The $CO_2$ profile and analytical model for the Pioneer Venus Large Probe neutral mass spectrometer. Icarus 392, 115374. https://doi.org/10.1016/j.icarus.2022.115374.

Mogul, R., Zolotov, M.Y., Way, M.J., Limaye, S., 2025. Re-analysis of Pioneer Venus data: Water, iron sulfate, and sulfuric acid are major components in Venus' aerosols. J. Geophys. Res. Planets 130, e2024JE008582. https://doi.org/10.1029/2024JE008582.

Parikh, S., Limbachiya, C., Joshipura, K., 2022. Calculations of total ionization cross-sections for electron impact on $H_2SO_4$. Proc. Internation. Conf. Atomic Mol. Opt. Nano Physics Applic. 529–534. https://doi.org/10.1007/978-981-16-7691-8_49.

Petrova, E.V., 2018. Glory on Venus and selection among the unknown UV absorbers. Icarus 306, 163–170. https://doi.org/10.1016/j.icarus.2018.02.016.

Petryanov, I., Andreichikov, B., Korchuganov, B., Ovsyankin, E., Ogorodnikov, B., Skitovich, V., Khristianov, V., 1981. Iron in the clouds of Venus. Soviet Phys. Doklady 260, 834–836.

Pieters, C., Head, J., Pratt, S., Patterson, W., Garvin, J., Barsukov, V., Basilevsky, A., Khodakovsky, I., Selivanov, A., Panfilov, A., 1986. The color of the surface of Venus. Science 234, 1379–1383. https://doi.org/10.1126/science.234.4782.1379.

Porshnev, N., Mukhin, L., Gel'man, B., Nenarokov, D., Rotin, V., Dyachkov, A., Bondarev, V., 1987. Gas chromatographic analysis of the products of thermal reactions of Venus cloud aerosols on the Vega-1 and 2 automated interplanetary probes. Cosm. Res. 2, 549–553.

Renggli, C.J., King, P.L., Henley, R.W., Norman, M.D., 2017. Volcanic gas composition, metal dispersion and deposition during explosive volcanic eruptions on the Moon. Geochim. Cosmochim. Acta 206, 296–311. https://www.sciencedirect.com/science/article/pii/S0016703717301503.

Robie, R.A., Hemingway, B.S., 1995. Thermodynamic Properties of Minerals and Related Substances at 298.15 K and 1 bar (105 Pascals) Pressure and at Higher Temperatures, U.S. Geological Survey Bulletin 2131. US Government Printing Office, Washington.

Seiff, A., Schofield, J., Kliore, A., Taylor, F., Limaye, S., Revercomb, H., Sromovsky, L., Kerzhanovich, V., Moroz, V., Marov, M.Y., 1985. Models of the structure of the atmosphere of Venus from the surface to 100 kilometers altitude. Adv. Space Res. 5, 3–58. https://doi.org/10.1016/0273-1177(85)90197-8.

Snow, K.B., Thomas, T.F., 1990. Mass spectrum, ionization potential, and appearance potentials for fragment ions of sulfuric acid vapor. Int. J. Mass Spectrom. Ion. Process. 96, 49–68. https://doi.org/10.1016/0168-1176(90)80041-Z.

Šolc, Z., Trojan, M., Brandova, D., Kuchler, M., 1988. A study of thermal preparation of iron (III) pigments by means of thermal analysis methods. J. Therm. Analysis Calorim. 33, 463–469. https://doi.org/10.1007/bf01913924.

Spratt, H., Rintoul, L., Avdeev, M., Martens, W., 2014. The thermal decomposition of hydronium jarosite and ammoniojarosite. J. Therm. Analysis Calorim. 115, 101–109. https://doi.org/10.1007/s10973-013-3213-1.

Surkov, Y.A., Ivanova, V., Pudov, A., Volkov, V., Sheretov, E., Kolotilin, B., Safonov, M., Thomas, R., Lespagnol, J., Hauser, A., 1986. VEGA-1 mass spectrometry of Venus cloud aerosols-preliminary results. Sov. Astron. Let. 12, 44.

Wallace, W.E., 2020. Mass Spectra" in NIST Chemistry WebBook, NIST Standard Reference Database Number 69. National Institute of Standards and Technology, Gaithersburg MD, 20899. https://doi.org/10.18434/T4D303.

Woitke, P., Scherf, M., Helling, C., Rimmer, P., Ferus, M., Lammer, H., Weichbold, F., Němečková, K., Eminger, P., Kačina, J., Constantinou, T., 2025. Prediction of sulfate hazes in the lower venus atmosphere. Planet. Sci. 6, 225. https://doi.org/10.3847/PSJ/ae00bf.

Xu, W., Parise, J.B., Hanson, J., 2010. $(H_3O)Fe(SO_4)_2$ formed by dehydrating rhomboclase and its potential existence on Mars. Am. Mineralog. 95, 1408–1412. https://doi.org/10.2138/am.2010.3470.

Yan, B., Li, L., Yu, Q., Hang, W., He, J., Huang, B., 2010. High irradiance laser ionization mass spectrometry for direct speciation of iron oxides. J. Am. Soc. Mass Spectrom. 21, 1227–1234. https://doi.org/10.1016/j.jasms.2010.02.030.

Zasova, L., Krasnopolsky, V., Moroz, V., 1981. Vertical distribution of $SO_2$ in upper cloud layer of Venus and origin of UV-absorption. Adv. Space Res. 1, 13–16. https://doi.org/10.1016/0273-1177(81)90213-1.

Zolotov, M., 2026. Sulfur on Venus: atmospheric, surface, and interior processes. In: Harlov, D.E., Pokrovski, G.S. (Eds.), The Role of Sulfur in Planetary Processes: From Cores to Atmospheres. Springer Geochemistry, New York, pp. 113–1314. https://doi.org/10.1007/978-3-032-07705-9_16.

Zolotov, M.Y., 2018. Gas–solid interactions on Venus and other solar system bodies. Rev. Mineral. Geochem. 84, 351–392. https://doi.org/10.2138/rmg.2018.84.10.

Zolotov, M.Y., 2021. Iron salts and oxides in the jistory of the surface-atmosphere-cloud system on Venus. Lunar Planet. Sci. Conf. 2615. https://www.hou.usra.edu/meetings/lpsc2021/pdf/2615.pdf.

Zolotov, M.Y., Mogul, R., Limaye, S.S., Way, M.J., Garvin, J.B., 2023. Venus cloud composition suggested from the pioneer Venus large probe neutral mass spectrometer data. Lunar Planet. Sci. Conf. 2806. https://www.hou.usra.edu/meetings/lpsc2023/pdf/2880.pdf.